\documentclass[aps,prd,showpacs,preprintnumbers,twelvepoint]{revtex4}
\usepackage{epsfig}
\usepackage{latexsym,amsmath,amssymb,amstext}

\newcommand\bef{\begin{figure}}
\newcommand\eef{\end{figure}}
\newcommand\beq{\begin{equation}}
\newcommand\eeq{\end{equation}}
\newcommand\beqa{\begin{eqnarray}}
\newcommand\eeqa{\end{eqnarray}}
\newcommand\bet{\begin{table}}
\newcommand\eet{\end{table}}

\newcommand\fgn[1]{Figure \ref{fig:#1}}
\newcommand\eqn[1]{eq.\ (\ref{#1})}
\newcommand\scn[1]{Section \ref{sec:#1}}
\newcommand\tbn[1]{Table \ref{table:#1}}

\newcommand\cdof{\chi^2/\mathrm{DOF}}
\newcommand\ie{{\sl i.e.\/}}
\newcommand\tr{\mathrm{Tr}\;}
\newcommand\x{\mathbf{x}}

\newcommand\etal{{\sl et al.\/}}
\newcommand\jhep{{\sl J.\ H.\ E.\ P.\/}\ }
\newcommand\np{{\sl Nucl.\ Phys.\/}\ }
\newcommand\pos{{\sl PoS\/}\ }
\newcommand\pr{{\sl Phys.\ Rev.\/}\ }
\newcommand\prlt{{\sl Phys.\ Rev.\ Lett.\/}\ }

\begin{document}
\title{Quasi-static probes of the QCD plasma}
\author{Debasish\ \surname{Banerjee}}
\email{debasish@theory.tifr.res.in}
\affiliation{Department of Theoretical Physics, Tata Institute of Fundamental
         Research,\\ Homi Bhabha Road, Mumbai 400005, India.}
\author{Rajiv\ V.\ \surname{Gavai}}
\email{gavai@tifr.res.in}
\affiliation{Department of Theoretical Physics, Tata Institute of Fundamental
         Research,\\ Homi Bhabha Road, Mumbai 400005, India.}
\author{Sourendu\ \surname{Gupta}}
\email{sgupta@theory.tifr.res.in}
\affiliation{Department of Theoretical Physics, Tata Institute of Fundamental
         Research,\\ Homi Bhabha Road, Mumbai 400005, India.}
\begin{abstract}
Screening correlators and masses were studied at finite temperature in
QCD with two flavours of dynamical staggered quarks on a lattice. The
spectrum of screening masses show a hierarchical approach to chiral
symmetry restoration. Control of explicit chiral symmetry breaking through
the quark mass was shown to be an important step to understanding this
phenomenon. No sign of decays was found in the finite temperature scalar
meson-like correlators in the confined phase.

\end{abstract}
\preprint{TIFR/TH/11-03}
\maketitle

\section{Introduction}\label{sec:intro}

High temperature QCD is the subject of a long-running experimental program
of heavy-ion collisions in several experimental facilities, including the
Relativistic Heavy Ion Collider (RHIC) at BNL, the Large Hadron Collider
(LHC) at CERN, and planned programs in GSI and Dubna. They aim to collide
together heavy-ions at high energies and produce fireballs of strongly
interacting matter, whose study may then yield information about the
phases and properties of such matter. Lattice computations are the primary
source of theoretical information on such matter. The aim of this study
is to examine the response of this matter to quasi-static perturbations.

It is known that the generic response of matter to such perturbations is
to screen them. The knowledge of screening lengths, $\xi$, is one of the
most basic pieces of microscopic information we can have about the system.
Of particular interest is the longest screening length, $\xi_0$, or its
inverse, the smallest screening mass, $\mu_0=1/\xi_0$ \cite{DeTarKogut}.
When $\mu_0=0$, the effective long-distance theory of matter (at, or
close to, equilibrium) requires us to take into account these unscreened
perturbations. Such is the case in a QED plasma, where magnetic fields
are not screened. Not only are equilibrium properties of a plasma strongly
influenced by this, but also the off-equilibrium long-distance theory
changes to magneto hydrodynamics rather than Navier-Stokes hydrodynamics,
as it is for other fluids.

For the QCD plasma with non-vanishing quark masses, one knows that all
fields are screened and the long-distance theory will be hydrodynamics
coupled to the diffusion of conserved charges. Nevertheless, the
study of screening masses is of practical importance.  If the smallest
dimension of the fireball produced in heavy-ion collisions is $\ell$,
one expects thermodynamic properties to manifest themselves only when
$\ell\mu_0\gg1$. Furthermore, since both these quantities are functions
of the temperature $T$, it is possible for the fireball to drop into,
and out of, equilibrium at different temperatures.  In this way,
quasi-static properties such as the screening masses may put bounds
on truly dynamical quantities such as the thermalization and freezeout
times in heavy-ion collisions.

The study of screening properties in a plasma has a long history. In
the glue sector, the Debye screening length has been the object
of many studies and now seems to be quantitatively understood,
both in non-perturbative lattice studies \cite{debye} and in weak
coupling theory at high temperatures \cite{helsinki}. Screening
in other quantum number channels in the glue sector has also been
studied \cite{glueball}. Screening in colour singlet channels due
to quark bilinear (meson-like) and trilinear (baryon-like) currents
\cite{DeTarKogut} was understood as the first signal of deconfinement
above the chiral symmetry restoring temperature in QCD with dynamical
quarks \cite{mtc}.  Analyticity arguments relate these hadron-like
screening masses in the low-temperature confined and chiral symmetry
broken phase to the (pole) masses and properties of the hadrons. This
has implications for models of heavy-ion collisions such as the hadron
resonance gas model.

One more application is to the viability of resummation of the weak
coupling series at high temperature using dimensional reduction.  This is
possible only if the lowest screening mass belongs to the glue sector
\cite{rvgsg}. It turns out that dimensional reduction does not work just
above the chiral cross over temperature, $T_c$.

Hadron-like screening masses have been studied extensively
\cite{catchall,wissel}. In QCD with light dynamical quarks they have
been studied before using 2 flavours of staggered quarks \cite{rvgsgpm}
and with 2+1 flavours of p4 improved quarks \cite{hotqcd}. They have
been studied also with overlap valence quarks and staggered sea quarks
\cite{rvgsgrl}. In all these studies the renormalized light quark masses
are almost equal, and nearly physical.

In this work, we extend previous studies through the analysis of
meson-like spatial correlation functions in 2-flavour QCD with staggered
quarks. This brings the state of the art for dynamical staggered quarks
into the regime of lattice spacings already reached using quenched
overlap quarks.  The organization of our paper is the following:
In \scn{tdet} we discuss operators selected for our analysis and the
technical details of our fitting methodology.  We investigate chiral
symmetry restoration through the correlation functions in \scn{csb}.
Our results on the screening spectrum are presented in \scn{scr} along
with a finite volume analysis.  Finally we summarize the main details
in \scn{summ}.  A technical point about the covariances of measurements
of correlators is dealt with in Appendix \ref{sec:fitstat}.

\section{Configurations, measurements and analysis}\label{sec:tdet}

A large part of this study uses decorrelated gauge configurations
described in  \cite{rvgsgNt6}. Two light flavours of staggered quarks were
used with the bare quark mass tuned so as to give $m_\pi\simeq230$ MeV at
zero temperature. The lattice spacing was $a=1/(6T)$, \ie, $N_t=6$. The
extraction of the temperature scale was explained in \cite{rvgsgNt6};
we note that $T_c$ was identified there through the peak of the Polyakov
loop susceptibility. The lattice volumes, $V=(a N_s)^3$ were set using
$12\le N_s\le24$, \ie, the aspect ratio $\zeta=T\sqrt[3]V=N_s/N_t$ between
2 and 4. This provided basic control of finite volume effects; most of
our results are reported for the largest volumes, $\zeta=4$. In addition,
we performed a detailed finite volume scaling study at $T=0.94T_c$. For
this we generated configurations with $4/3\le\zeta\le5$, with all other
parameters fixed as before. We measured autocorrelation times as before,
and used at least 50 decorrelated configurations for our measurements.

The meson screening correlation functions projected to zero momentum are---
\beq
 C^\gamma_z = \frac1{\cal V}\left\langle\sum_{\x}\tr\left[G(\x,z)
     G^{\dagger}(\x,z)\right] \phi_\gamma(\x)\right\rangle
\label{meas}\eeq
where $\x$ stands for sites labelled by the triplet $(x,y,t)$, the
number of terms in the sum is the same as the volume of such a slice,
${\cal V}=N_x N_yN_t$, $G(\x,z)$ is the inverse of the Dirac operator,
\ie, the quark propagator from the origin to the point $(\x,z)$, the
angular brackets denote an average over gauge field configurations with
the correct weight, and the staggered phase factors $\phi_\gamma(\x)$
pick out the quantum numbers, $\gamma$, of the meson under study.

In this work we have taken all eight possible local staggered phases. At
$T=0$ they would correspond to the flavour non-singlet scalar (S)
(corresponding to the $a_0$ meson), the Goldstone pion (PS), and three
components each of the local vector meson (V) and the axial vector (AV).
Symmetry operations of the spatial slice interchange the components of
the V and AV, so the three components are expected to be identical after
averaging over gauge configurations.

Since we measure spatial direction correlators at finite temperature,
the symmetries of the $(x,y,t)$ slice orthogonal to the direction of
propagation are not the same as they would be in the corresponding zero
temperature computation \cite{sg1999}.  The S/PS operators both lie
in the trivial representation, called the $A_1^{++}$, of the spatial
direction transfer matrix.  The sum of the $x$ and $y$ polarizations
of the V/AV, and, separately, the $t$ polarization, also lie in the
$A_1^{++}$ representation. These six different kinds of $A_1^{++}$
operators do not mix under the symmetries of the $(x,y,t)$ slice, and
hence we need separate notations for them. For the S/PS correlators it is
economical to carry on the $T=0$ notation. For the sum of the $x$ and $y$
polarizations of the V we use the notation Vs (and AVs for the sum in the
AV sector) and for the $t$ polarizations we use the notation Vt and AVt.
The difference of the $x$ and $y$ polarizations of the V/AV lie in a
non-trivial representation called the $B_1^{++}$. We use the notation
VB and AVB for these. These particular realizations of the $B_1^{++}$
correlator have earlier been seen to vanish \cite{rvgsg1999}.

We will also have occasion to use the S and PS susceptibilities \cite{sgsus}
defined as
\beq
   \chi_{PS} = \sum_z C^{PS}_z,\qquad{\rm and}\qquad
   \chi_S = \sum_z (-1)^z C^{PS}_z.
\label{suscep}\eeq
The construction uses a fact that we demonstrate later: at high temperatures
the S/PS correlators are essentially dominated by a single parity state.

\bef[!tbh]
\begin{center}
\includegraphics[scale=0.5,angle=270]{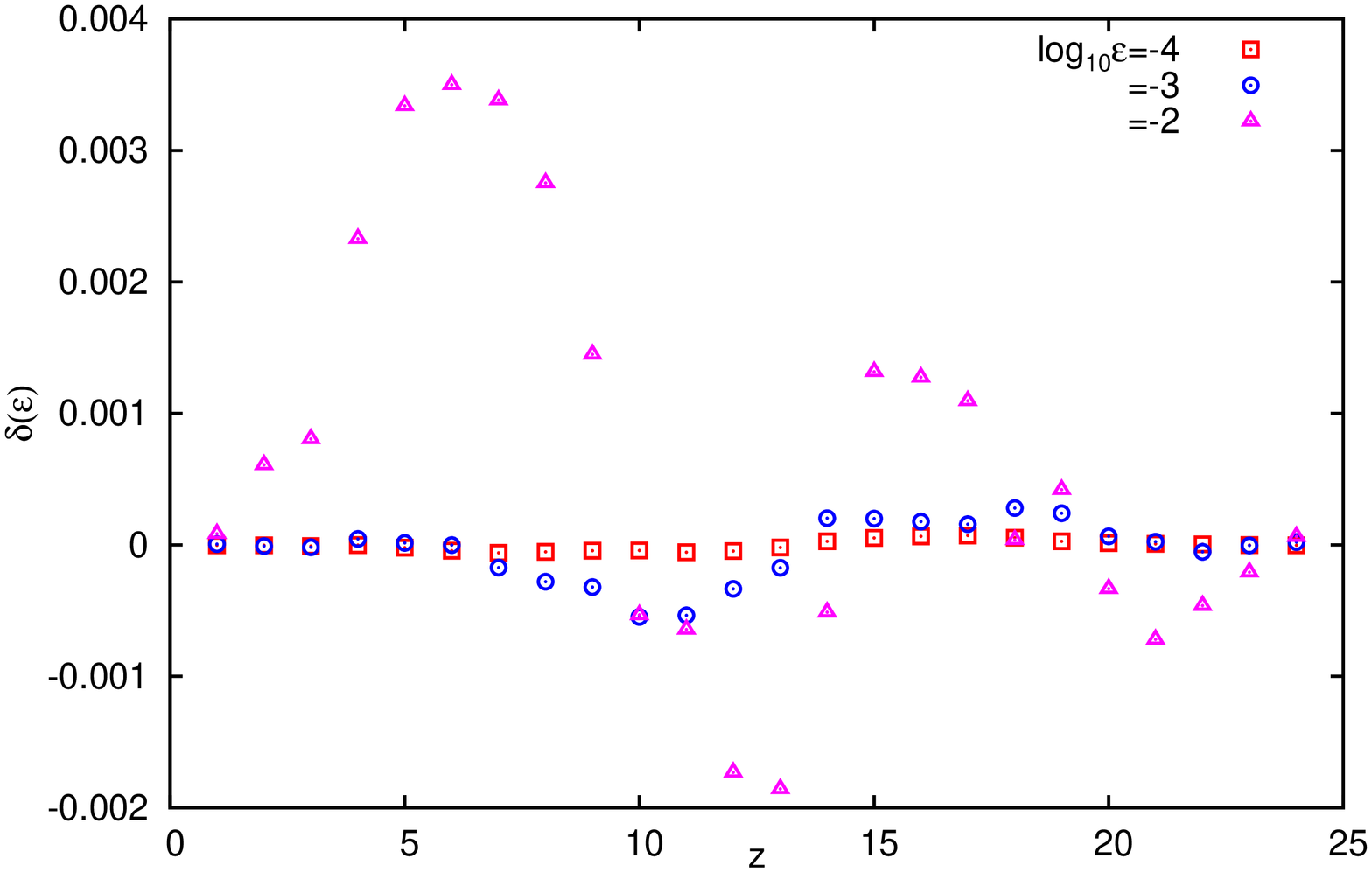}
\end{center}
\caption{$\delta(\epsilon)$, see \eqn{rat}, as a function of the
separation $z$ for varying $\epsilon$, with a fixed value of
$\epsilon'=10^{-5}$.}
\label{fig:convg}\eef

The inversion of the Dirac operator was done using a conjugate gradient
(CG) algorithm, as usual. The tolerance for stopping was chosen
such that the residual vector had squared norm less than $\epsilon N_tN_s^3$.
We investigated whether we had an acceptable stopping criterion by monitoring
\beq
 \delta(\epsilon) = 1-\frac{C^{PS}[\epsilon]}{C^{PS}[\epsilon']}
\label{rat}\eeq
where $C^{PS}[\epsilon]$ is the PS correlation function obtained when
the stopping tolerance parameter is $\epsilon$.  We chose a fixed
$\epsilon'=10^{-5}$.  In \fgn{convg} we show $\delta$ computed on a
randomly chosen test configuration at $T=0.94T_c$ with $\zeta=4$. The
configuration to configuration variance of $C^{PS}$ is about 2--5\%
of the expectation value, so keeping $\delta<0.01$ suffices. Clearly,
the errors converge fast, and our choice of $\epsilon=10^{-5}$ is seen
to be more than sufficient.

Each staggered correlation function may contain contributions
from two parity partners, and can be parametrized through the
doubled-parity fit
\beq
C(z) = A_1(\mathrm{e}^{-\mu_1 z} + \mathrm{e}^{-\mu_1(N_z-z)})
      + (-1)^z A_2(\mathrm{e}^{-\mu_2 z} + \mathrm{e}^{-\mu_2(N_z-z)}),
\label{2masfit}\eeq
where $\mu_1$ and $\mu_2$ are the screening masses of the lightest natural
parity meson appropriate for the operator used and its opposite 
parity partner.

Since measurements of the correlation function at different distances,
$z$, are made using the same gauge configurations, they are correlated,
and the fit must take care of these correlations. Therefore, we used
the definition of $\chi^2$
\beq
\chi^2 = \sum_{zz'} 
  \left[C_z-C(z)\right]\Sigma_{zz'}^{-1}\left[C_{z'}-C(z')\right].
\eeq
Here $z$ is the spatial separation, $C_z$ are the measured expectation
values of \eqn{meas}, the function $C(z)$ is the 2-mass form of
\eqn{2masfit}, and $\Sigma_{zz'}$ is the covariance of $C_z$ and
$C_{z'}$. When $\Sigma_{zz'}$ is diagonal, the definition reduces to the
more familiar one. In actuality, the correlation coefficients are fairly
high, so the matrix $\Sigma_{zz'}$ is nearly singular. The inversion
was done in Mathematica to an accuracy of ${\cal O}(10^{-10})$. The errors
in the inversion were therefore negligible compared to the statistical
errors in the measurements, $\sigma_z$, which were of the order of a
few percent. See Appendix \ref{sec:fitstat} for further discussion of
this procedure.

A check on the consistency of the results obtained from fits is to use
local masses. Due to the even-odd oscillations for staggered
fermions, we used the definition of \cite{rvgsgpm}---
\beq
 \frac{C_{z+1}}{C_{z-1}} =
 \frac{\cosh[-m(z)(z + 1 - N_z/2)]} {\cosh[-m(z)(z - 1  - N_z/2)]}.
\label{localm}\eeq
Given the measurement on the left, the effective mass, $m(z)$, can be
extracted by solving the equation and errors estimated by jack-knife.
This differs from a procedure where successive time slices are used for
the modified correlator $(-1)^zC_z$ \cite{hotqcd}.

In the chiral symmetry broken phase there is no particular relation
between $\mu_1$, $\mu_2$ and $A_1$, $A_2$ for different correlators. However,
when chiral symmetry is restored, the staggered phases give
\beq
  C^{PS}_z = (-1)^z C^S_z,\qquad 
  C^{AVs}_z = (-1)^z C^{Vs}_z,\qquad
  C^{AVt}_z = (-1)^z C^{Vt}_z.
\label{symm}\eeq
This implies the relations
\beq
   A_1^{Vs}=A_2^{AVs},\qquad \mu_1^{Vs}=\mu_2^{AVs}\quad
       \mathrm{and}\quad (Vs\leftrightarrow AVs),
\label{csrest}\eeq
and similarly for the Vt and AVt or the S and PS channels. These relations
are very easily demonstrated by using the projections
\beq
   C^{(\pm S)}_z = C^{PS}_z \pm (-1)^z C^S_z, \qquad
   C^{(\pm Vs)}_z = C^{Vs}_z \pm (-1)^z C^{AVs}_z, \qquad
   C^{(\pm Vt)}_z = C^{Vt}_z \pm (-1)^z C^{AVt}_z.
\label{chiralproj}\eeq
If the correlators $C^{(-\gamma)}_z$ vanish for all $z$ then chiral symmetry
is restored for the full spectrum of excitations.

\section{Thermal Effects and Approximate Chiral Symmetry Restoration}\label{sec:csb}

\bef[!tbh]
\begin{center}
\includegraphics[scale=0.70]{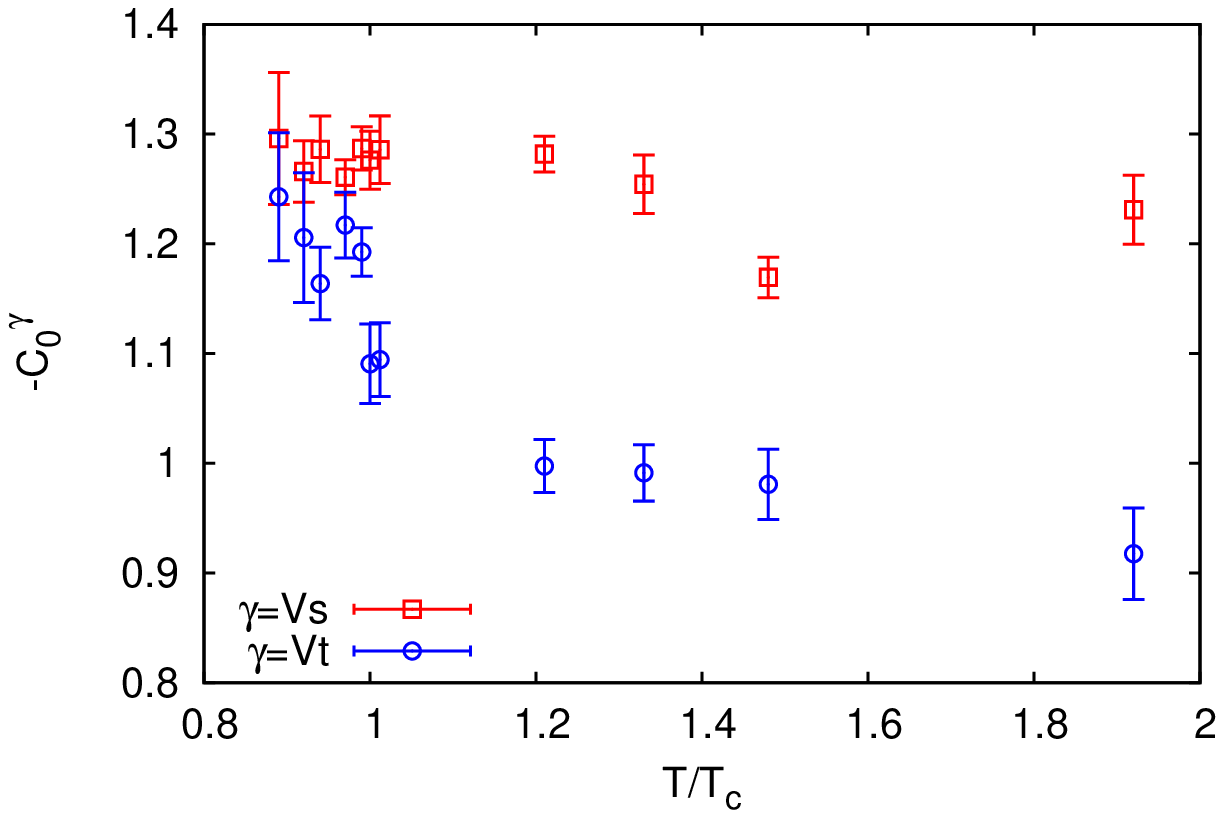}
\includegraphics[scale=0.70]{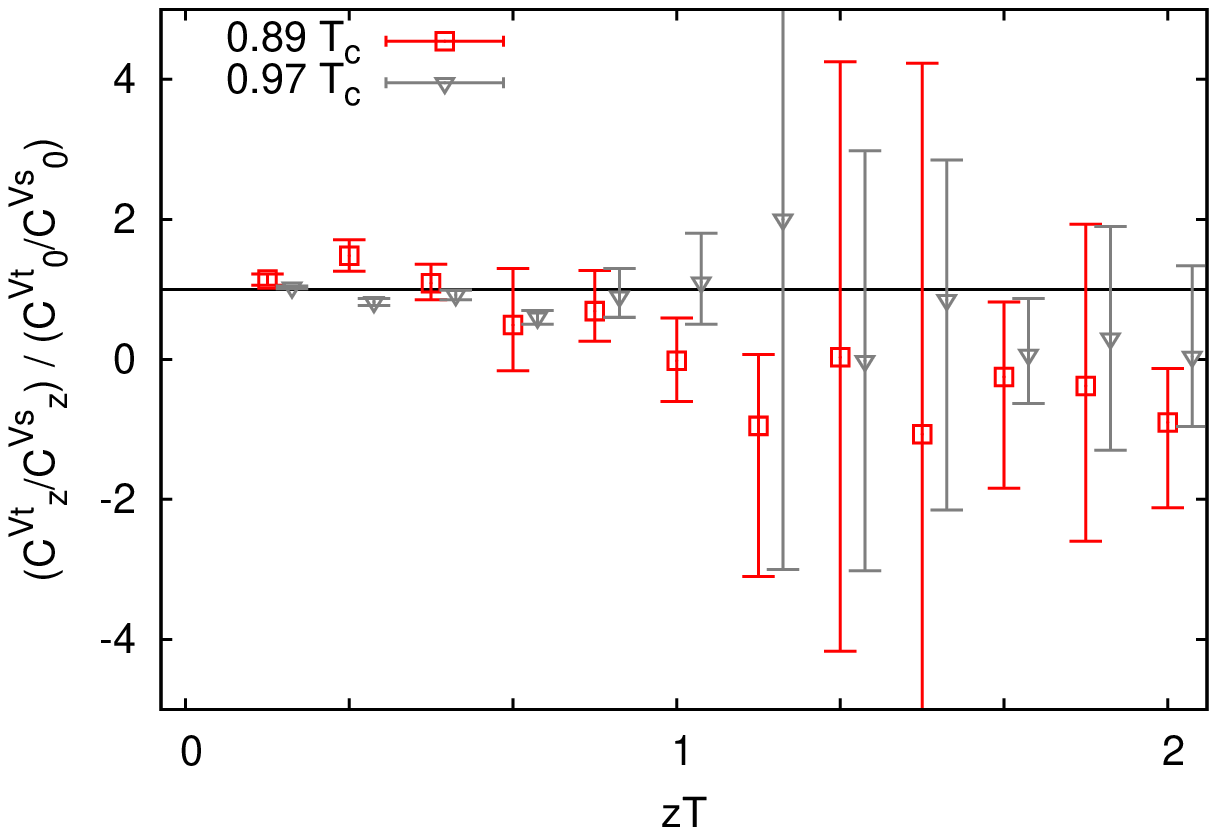}
\end{center}
\caption{The panel on the left shows Vs and Vt correlators at $z=0$ as
a function of $T/T_c$. Note the abrupt lifting of the degeneracy at $T_c$.
The panel on the right shows the ratio of the Vs and Vt correlators as a
function of $z$, normalized by their values at $z=0$. The data at $T=0.97
T_c$ is displaced slightly to the right for clarity.}
\label{fig:vector}\eef

The screening correlators at any non-zero temperature should be
decomposed according to the symmetry group of the finite temperature
slice. At sufficiently low temperature, however, one expects the Vt
and Vs correlators to be nearly equal, and the symmetries of the $T=0$
problem to be realized approximately. We investigated this by computing
the ratios of the Vt and Vs correlators (normalized to be unity at
$z=0$). The statistical analysis was performed using a bootstrap,
since the distribution of the ratio is not expected to be Gaussian
\cite{rvgsgNt6}. The results below $T_c$ are shown in \fgn{vector};
the ratio is consistent with unity at all $z$.  The normalization is
the value of the ratio of the correlators at $z=0$. In \fgn{vector}
we have plotted $C^{Vs}_0$ and $C^{Vt}_0$ as a function of $T$.
Below $T_c$ the two are equal within statistical errors. The two facts
taken together imply that the $T=0$ symmetries remain good until rather
close to $T_c$. Quite abruptly, just above $T_c$ this higher symmetry is
broken, and the symmetry of the finite temperature problem is obtained.
Similar results are obtained for the AVs and AVt. In view of this, in
most of our subsequent analysis, we will group the correlators below
$T_c$ into S, PS, V and AV.  For $T\ge T_c$ we will continue to use the
decomposition into S, PS, Vs, Vt, AVs and AVt.

\bef[!ptbh]
\begin{center}
\includegraphics[scale=0.82]{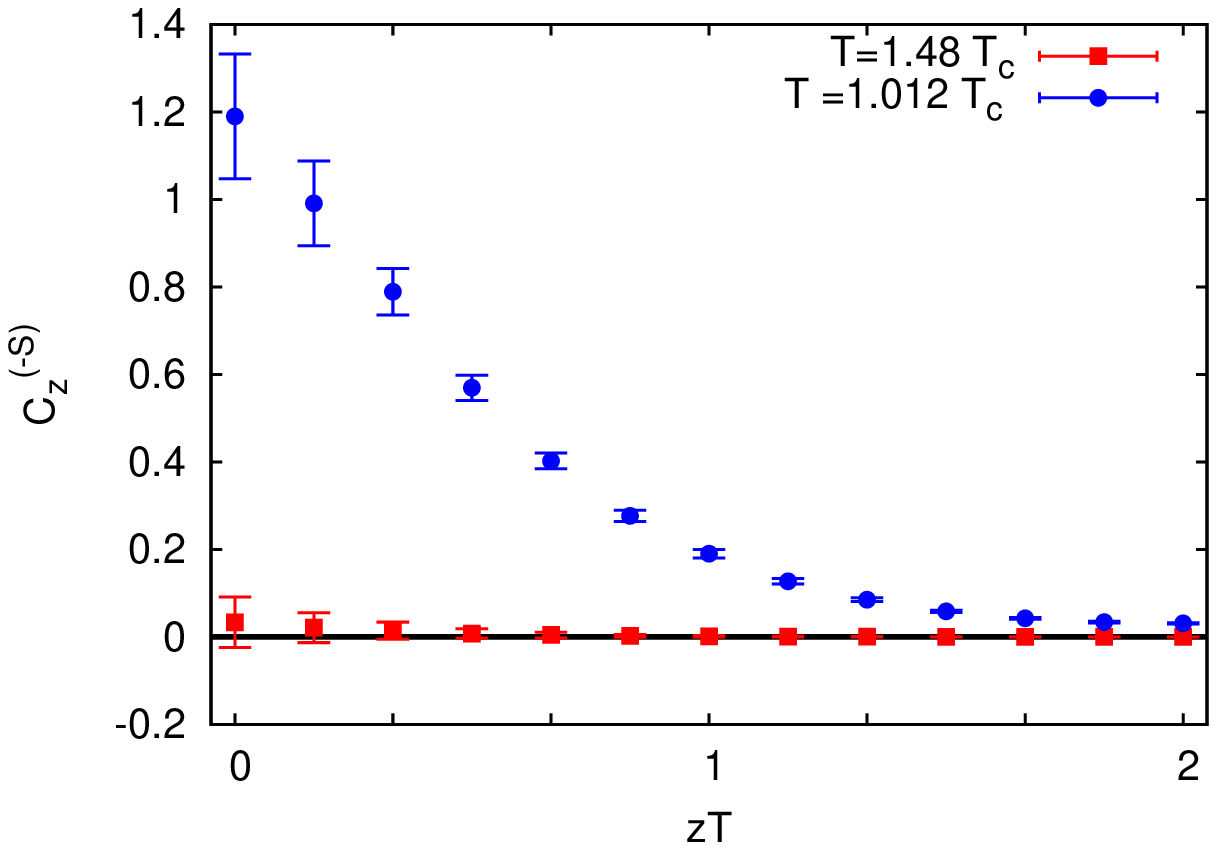}
\includegraphics[scale=0.82]{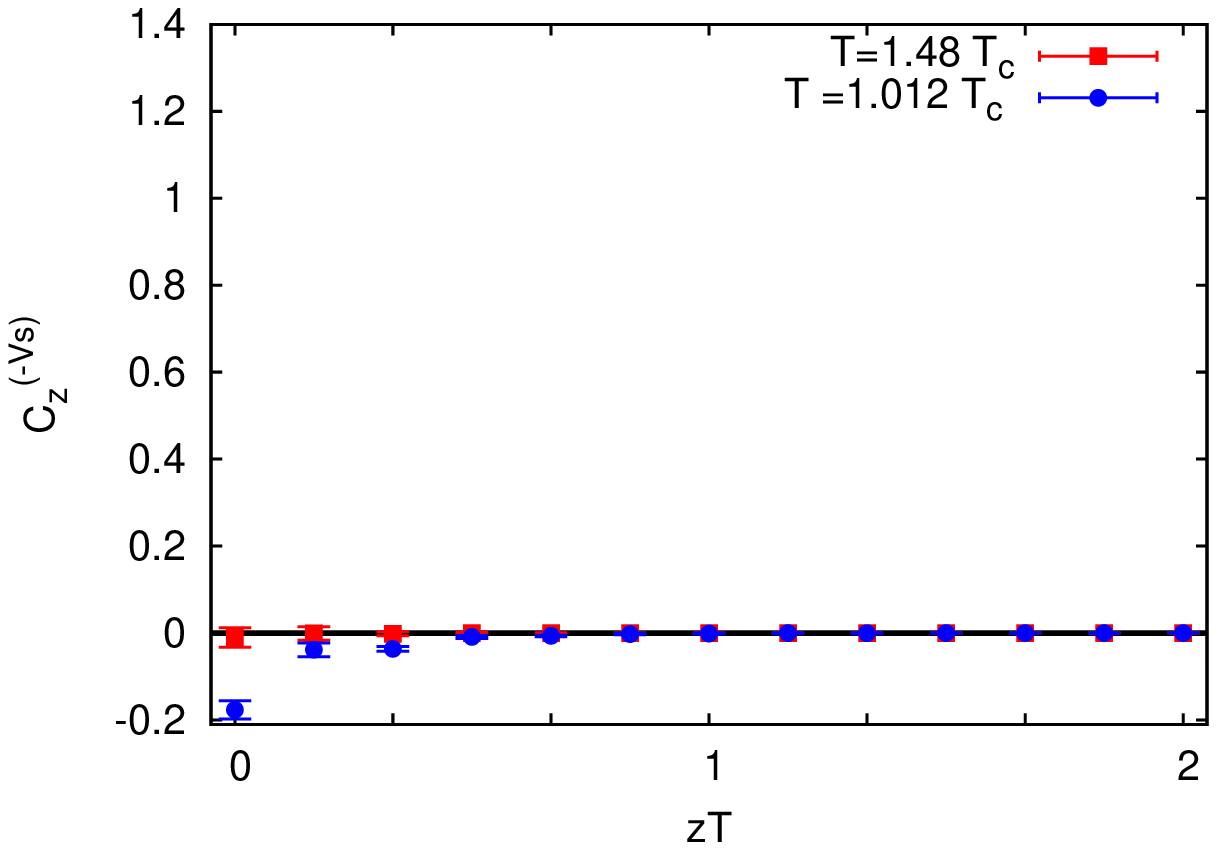}
\includegraphics[scale=0.82]{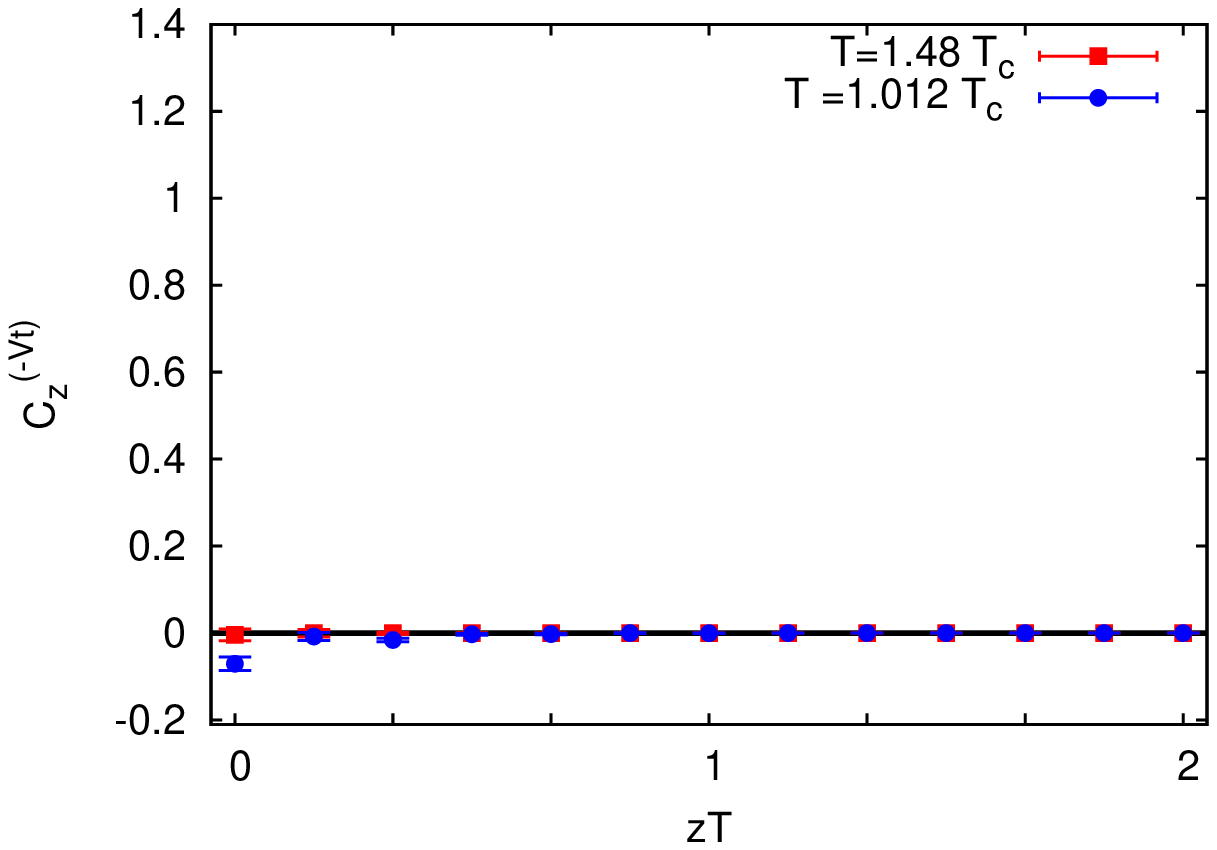}
\end{center}
\caption{The chiral projections $C^{(-S)}_z$, $C^{(-Vs)}_z$ and
$C^{(-Vt)}_z$ of \eqn{chiralproj} at two temperatures as computed
on lattices with $\zeta=4$. The Vt channel shows chiral symmetry
is close to being restored immediately above $T_c$.}
\label{fig:chproj3}\eef

The next question which we examine is whether the correlation functions
exhibit chiral symmetry restoration at any temperature. As discussed
earlier, the most straightforward way to examine this is to plot
$C^{(-\gamma)}_z$ at each $T$ and ask whether it is consistent with
zero at all $z$.  In \fgn{chproj3} we show these quantities at two
temperatures above $T_c$. From the correlator $C^{(-S)}_s$ we see that at
$T=1.48T_c$ the symmetry is clearly restored, whereas for $T=1.012T_c$
the symmetry is broken. The correlator $C^{(-Vs)}_z$ is consistent
with zero for $zT>1$, but at distances less than $1/T$ there is clear
chiral symmetry breaking close to $T_c$.  The correlator $V^{(-Vt)}_z$
most nearly exhibits chiral symmetry restoration immediately above $T_c$,
with only the value at $z=0$ being significantly non-zero.  At higher
temperatures all the $C^{(-\gamma)}_z$ are consistent with zero at all $z$,
thereby indicating chiral symmetry restoration.

\bet[!tbh]
\begin{center}
\begin{tabular}{|c|c|c|c|c|c|}
\hline
$T/T_c$ & $\chi^2_S$  &  $\chi^2_{Vs}$ & $\chi^2_{Vt}$ & $\chi^2_{VB}$ & $\chi^2_{AVB}$ \\
\hline
1.92   & 0.52 &   0.42  &   0.86 &  16.2   &  16.2    \\
1.48   & 9.62 &   1.33  &   0.77 &  10.2   &  11.5    \\
1.33   & 13.5 &   2.07  &   1.13 &  14.1   &  11.7    \\
1.21   & 441  &  27     &  15    &  19.8   &  8.8     \\
1.012  &1429  & 107     &  40    &   6.2   &  12.1    \\
1.00   & 878  & 125     &  56    &  19.0   &   9.1    \\
0.99   &3009  & 214     & 114    &  10.1   &  17.2    \\
0.97   &3013  & 440     & 212    &  14.0   &   9.2    \\
0.94   & 746  & 188     &  92    &  28     &  14.9    \\
0.92   & 936  & 348     & 141    &  15.0   &  14.5    \\
0.89   & 539  & 101     &  80    &  21.3   &  18.8    \\
\hline
\end{tabular}
\end{center}
\caption{This table lists the values of $\chi^2$ at different temperature
for tests of the hypotheses that various correlators vanish. The number
of degrees of freedom in all these cases is 12, since there are 13 independent
values of $z$ on the lattices with $\zeta=4$ with periodic boundary conditions.
In order to rule out the hypothesis that a correlator vanishes at the 99\% CL,
the value of $\chi^2$ should be more than 36.}
\label{table:zero}\eet

In order to extend this analysis to all temperatures it is useful
to introduce a less local quantity,
\beq
   \chi_\gamma^2 = \sum_{zz'} C^{(-\gamma)}_z \Sigma^{-1}_{zz'} C^{(-\gamma)}_{z'},
\label{csbmeasure}\eeq
where $\Sigma_{zz'}$ is the covariance matrix of the measurements of the
correlator at different distances. This is a measure of the likelihood
that the correlators at all $z$ are consistent with zero, and hence that
chiral symmetry is restored. Values of these variables are collected in
\tbn{zero}. Note that $\chi_{Vt}^2$ shows a distinct change at $T_c$,
although it is consistent with chiral symmetry restoration only at
$T=1.21T_c$. $\chi_{Vs}$ also shows a change at $T_c$, although it is
less dramatic. From \fgn{chproj3} it would appear that the change in
$\chi_{Vs}^2$ at $T_c$ is due to the long distance ($z>1/T$) correlation
function becoming consistent with zero, whereas the short distance
($z\le1/T$) part disappears only at larger $T$. $\chi_S^2$, on the other
hand, does not seem to undergo any significant change at $T_c$ and signals
chiral symmetry restoration only at $T=1.33T_c$. One sees the difference
in behaviour  in \fgn{chproj3}; the S/PS correlators, unlike the V/AV,
do not show any kind of effective long distance chiral symmetry restoration.
This spatial structure has not been noticed before, and could be worth
further investigation in future.

The late restoration of chiral symmetry breaking can be understood from
the fact that the non-vanishing quark mass provides explicit chiral
symmetry breaking. In the chiral limit, there is a phase transition at
$T_c$. In the high temperature phase there is, effectively, a single
scale, $T$, so the screening mass $\mu\propto T$. However, when there
is a non-vanishing bare quark mass, $m$, there is no phase transition at
$T_c$ but only a cross-over.  In the absence of a  phase transition, one
could have $\mu/T=f(m_\pi/T)$ where $m_\pi$ is the pion mass at $T=0$
(we have traded the bare quantity $m$ for a renormalized measure of
chiral symmetry breaking, $m_\pi$). At large $T$, when the argument of
the function becomes small, $f$ should go to a constant. On lowering $T$
from large values, non-constant behaviour should become visible when
the argument becomes of order unity, \ie, at $T/T_c\simeq m_\pi/T_c$.
Since our simulations are performed with $m_\pi\simeq230$ MeV and
$T_c\simeq 175$ MeV, this argument implies that explicit chiral symmetry
breaking should manifest itself up to $T\simeq1.35T_c$, which is what
we see. Such an argument would lead us to expect that in the real world
chiral symmetry breaking in screening masses should not be visible
above $T_c$. This may have some bearing on the relation between $T_c$
defined through susceptibilities of the deconfinement and chiral order
parameters \cite{tc}. An alternative explanation of the late restoration
of chiral symmetry is due to approximate restoration of the U$_{\mathrm
A}$(1) symmetry \cite{hotqcd}. A future computation with different sea quark
masses can easily distinguish between these two alternatives.

The projection of the local V and AV channel correlators on the $B_1^{++}$
channel is expected to vanish. This was demonstrated in \cite{rvgsg1999,
rvgsgpm} with lattice spacing $a=1/(4T)$. Here we investigate the
vanishing of these correlators at smaller lattice spacing using a
correlated $\chi^2$ definition similar to that above. In $\chi_{VB/AVB}^2$
the factors of $C_z^{-\gamma}$ in \eqn{csbmeasure} are replaced by the VB
or AVB correlator.  The results are collected in \tbn{zero} and show that
the VB and AVB correlators vanish.

\bef[!tbh]
\begin{center}\includegraphics{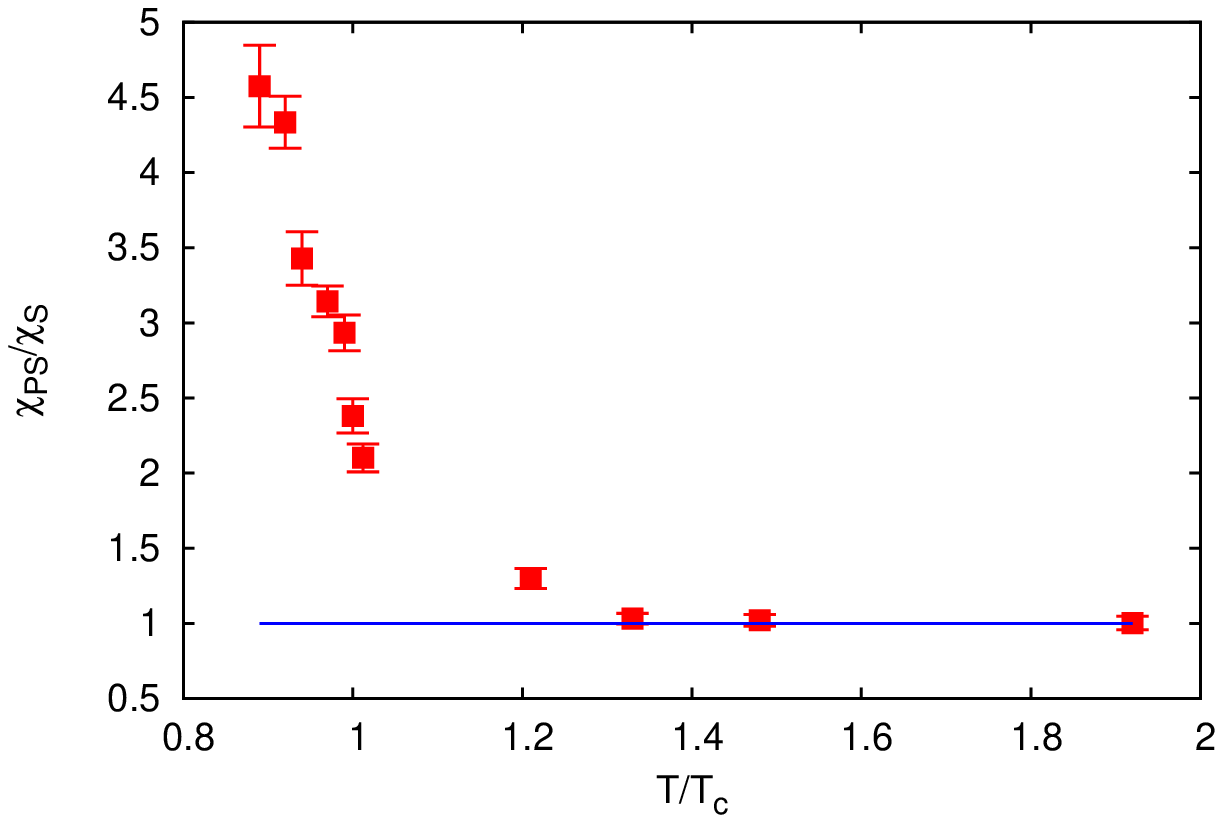}\end{center}
\caption{The ratio for susceptibilities of the pion to the scalar meson.}
\label{fig:sus1}\eef

In \fgn{sus1} we show the ratio of the PS and S susceptibilities
(see eq.\ \ref{suscep}). As expected, they become equal at $T=1.33T_c$,
which is the point where the two correlators begin to satisfy \eqn{symm}.
At lower temperatures $\chi_{PS}$ is larger, essentially because $\mu_{PS}$
is smaller than $\mu_S$.

\bef[!tbh]
\begin{center}
\includegraphics[scale=0.45]{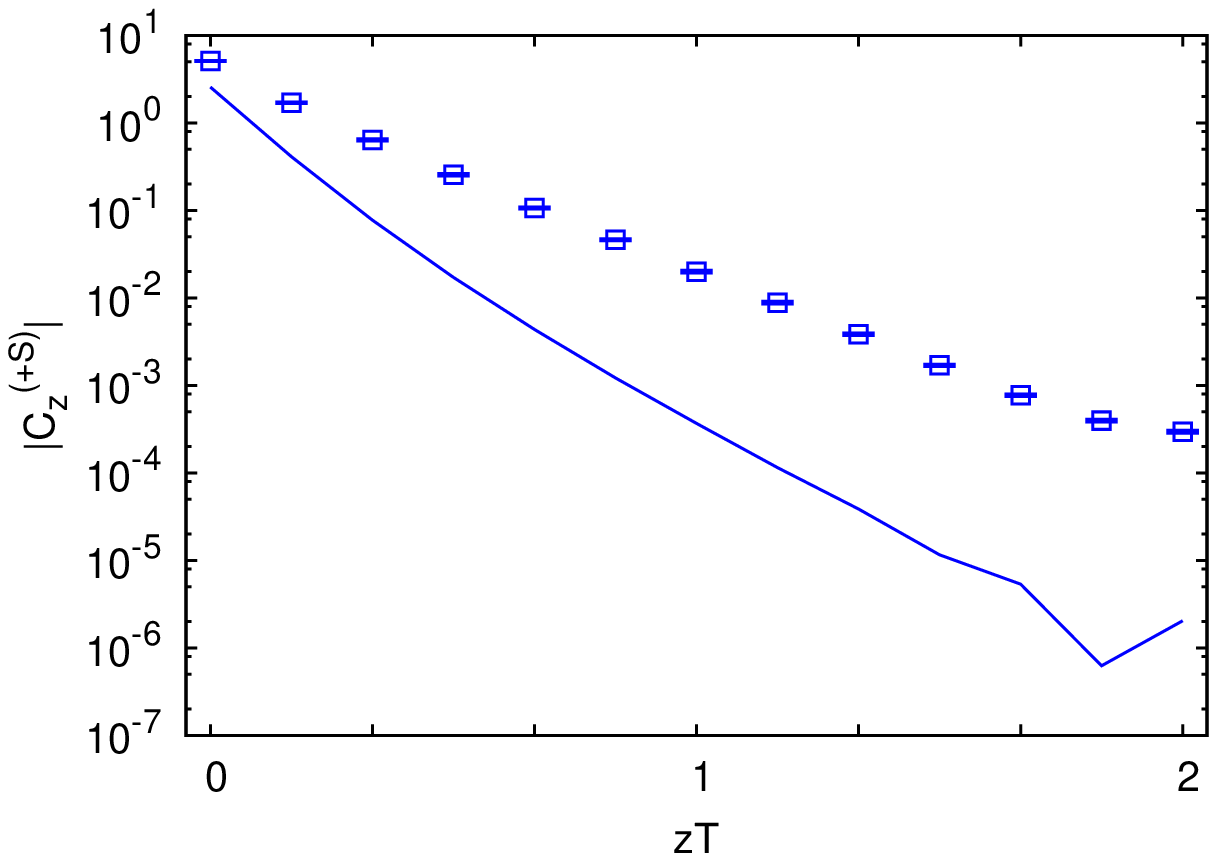}
\includegraphics[scale=0.45]{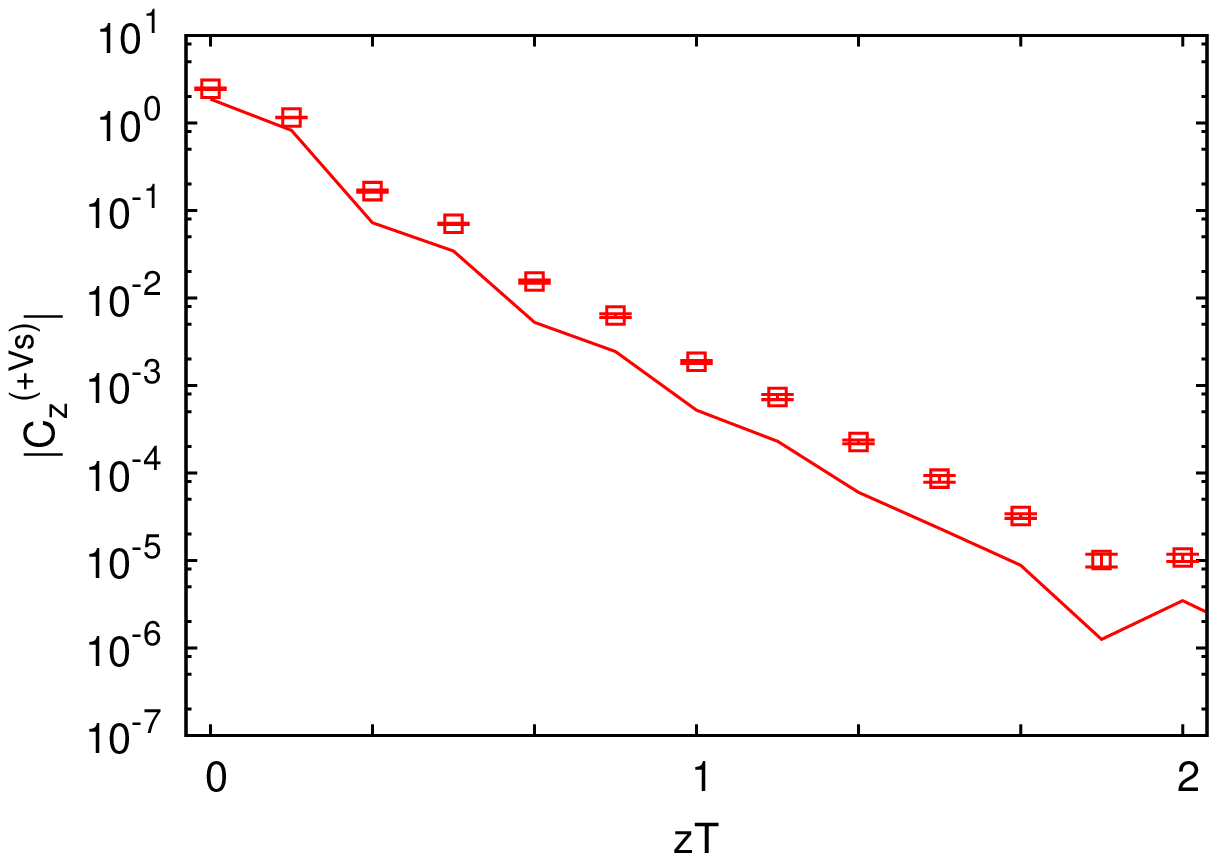}
\includegraphics[scale=0.45]{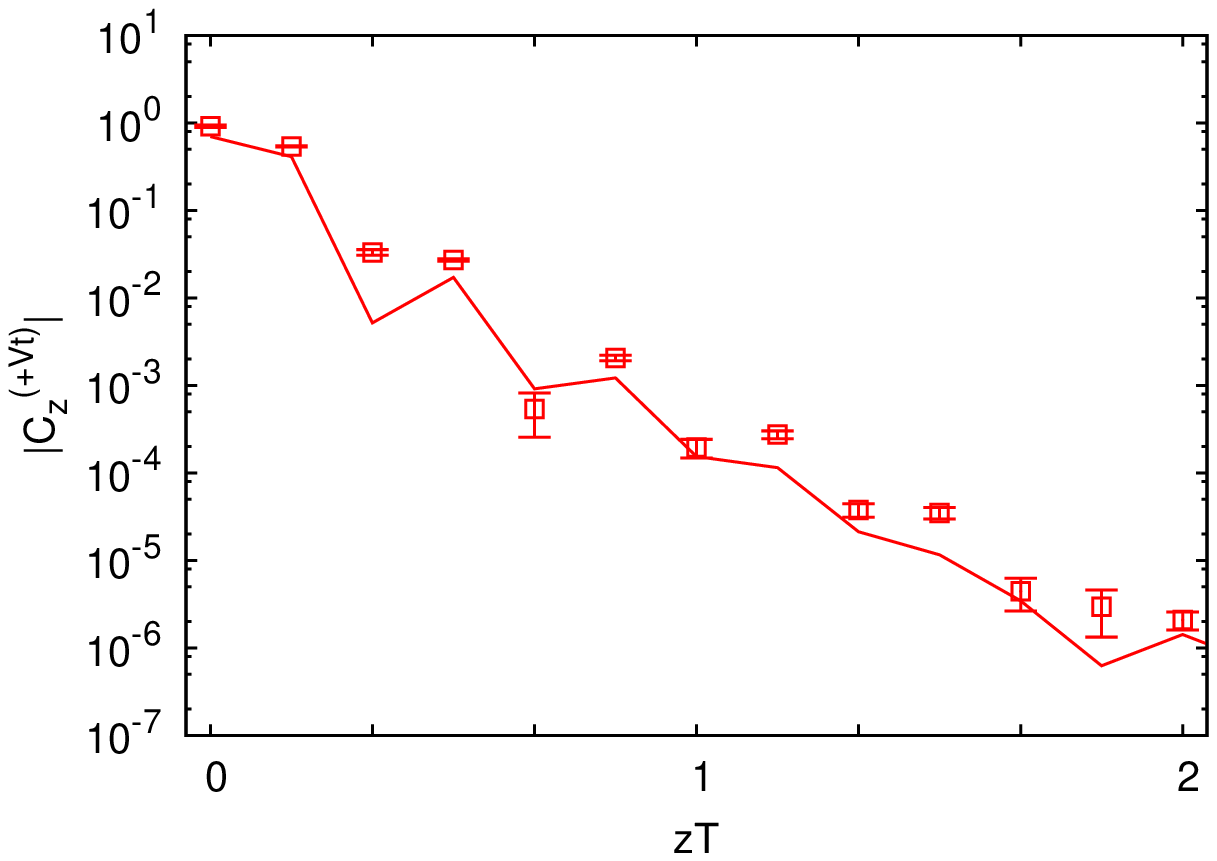}
\end{center}
\caption{The correlators $C^{(+\gamma)}_z$ obtained at $T=1.92T_c$ with bare
quark mass $m/T=0.1$. The lines show the result for free field theory
computed with the same quark mass.} 
\label{fig:chproj1}\eef

In the next sections we examine the other projection of the correlation
function, $C^{(+\gamma)}_z$.  This is non-zero at all temperatures. At
high enough temperature one might expect the whole correlation function
to be described in a weak-coupling theory. In \fgn{chproj1} we show the
correlators at $T=1.92T_c$.  One sees that the correlation function is far
from the free field theory result, especially the correlator $C^{(+S)}_z$;
indicating that even at this temperature the theory cannot be treated
as weakly interacting. This is consistent with previous observations
\cite{rvgsgpm,hotqcd,rvgsgrl}.

\section{Screening Masses}\label{sec:scr}

\bef[!tbh]
\begin{center}
\includegraphics[scale=0.7]{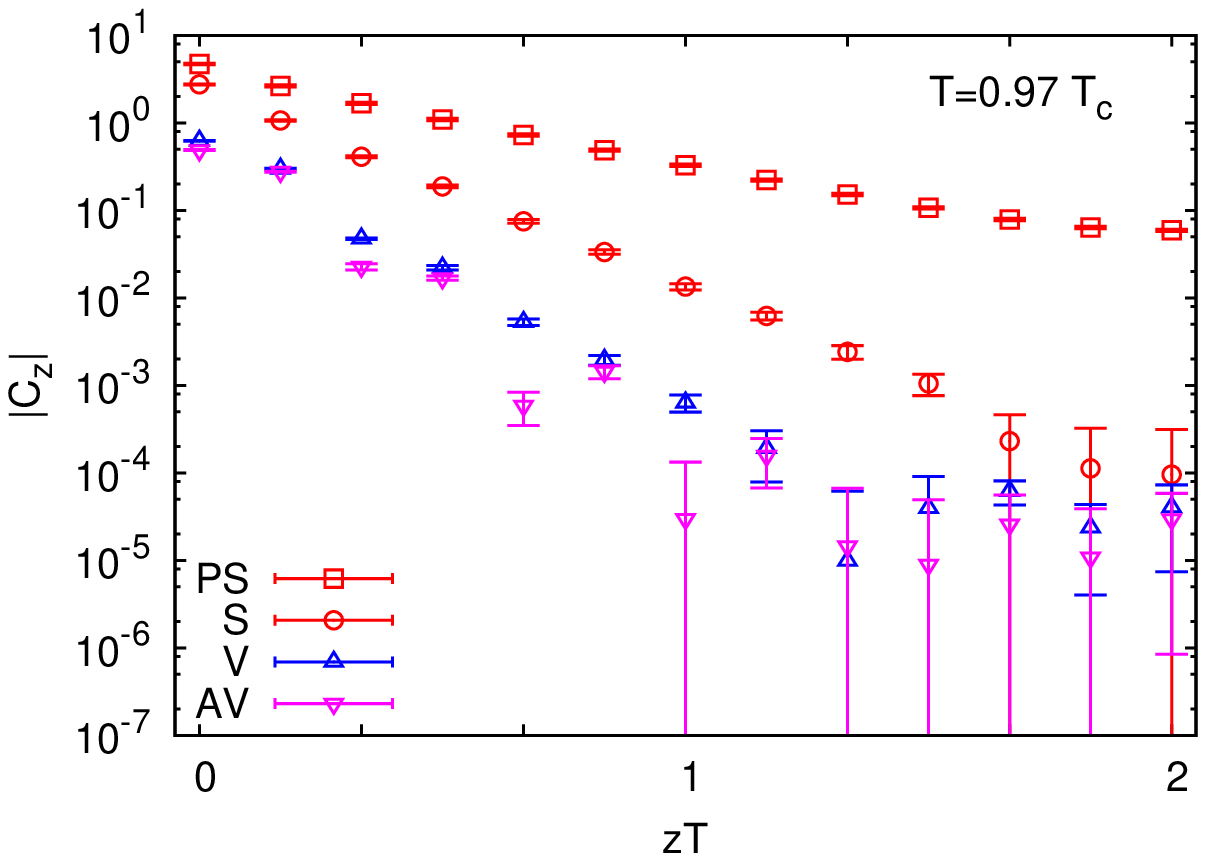}
\includegraphics[scale=0.7]{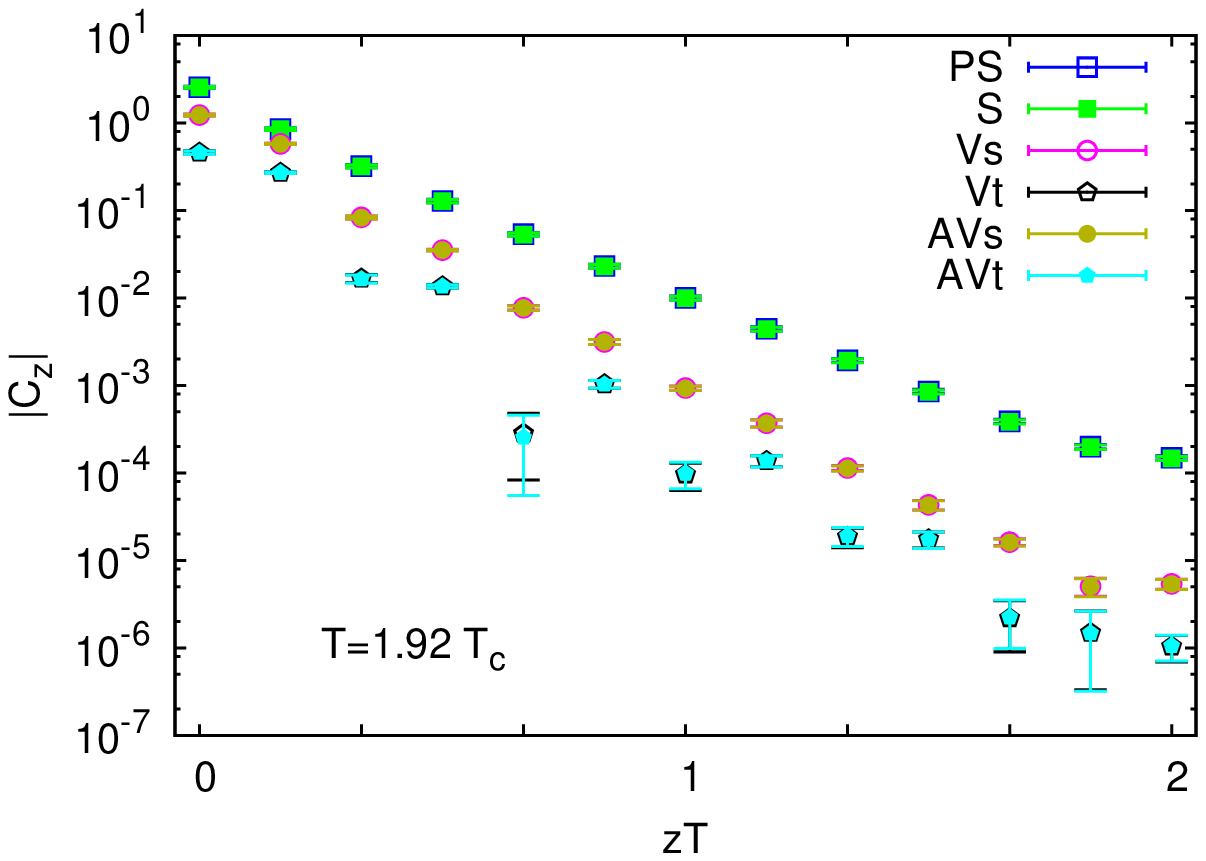}
\end{center}
\caption{The correlators for all the mesons at representative
temperatures in the two phases on the lattice with $\zeta=4$. Note
the absence of even-odd oscillations for the S and PS correlators.}
\label{fig:corrl}\eef

\bet[!tbh]
 \begin{center}
 \begin{tabular}{c||c|c|c|c||c|c|c}
 \hline
 &\multicolumn{4}{|c}{PS} &\multicolumn{3}{|c}{V} \\
 \hline
 &\multicolumn{7}{|c}{Uncorrelated single parity fit}\\
 \hline
 Range& 3--11 & 4--11 & 5--11 & & 4--9 & 4--10 & \\
  \hline
 $A$ & 3.61(7) & 3.59(6) & 3.59(6) & & & & \\
 $m$ & 0.401(2) & 0.400(2) & 0.400(2) & & & & \\
 $\cdof$ & 0.24/7 & 0.01/6 & 0.01/5 & & & & \\
 \hline
 &\multicolumn{7}{|c}{Correlated single parity fit}\\
\hline
$A$ & 3.64(18) & 3.59(16) & 3.58(16) & & 0.053(4) & 0.053(3) & \\
$m$ & 0.401(3) & 0.400(3) & 0.400(2) & & 0.484(7) & 0.480(8) & \\
$\cdof$ &  11.0/7 & 2.1/6 & 1.9/5   & & 9.5/4 & 9.8/5 & \\
 \hline
 &\multicolumn{7}{c}{Correlated doubled-parity fit}\\
 \hline
 & & 4--11 & 5--11 & 6--11   & 2--11 & 1--10 & 3--9 \\
 \hline
 $A_1$ & & 3.59(14)  & 3.59(12) & 3.58(11) & -0.67(4) & -0.70(3) & -0.55(5) \\
 $\mu_1$ & & 0.400(2)  & 0.400(2) & 0.400(2) & 1.18(3) & 1.20(3) & 1.14(3) \\
 $A_2$ & & 0.02(6)   & 0.04(12) & 3.1(*)   & 0.38(6) & 0.49(4) & 1.4(4) \\
 $\mu_2$ & & 0.9(*)  & 0.9(*) & 1.5(*)       & 1.62(8) & 1.71(5) & 1.98(11) \\
 $\cdof$ & & 1.60/4 & 1.58/3 & 1.28/2 & 6.55/6 & 6.56/6 & 2.20/3 \\
 \hline
\end{tabular}
\end{center}
\caption{Fits to PS and V Correlators at $T=0.97T_c$ on a lattice with
$\zeta=4$. An asterisk on a number indicates that the fit value is not
determined reliably.}
\label{table:massPS_b5.415}\eet

\fgn{corrl} displays all four screening correlators at two temperatures,
one each in the hadron and the plasma phase, \ie, at $T=0.97T_c$ and
$T=1.92T_c$.  The V/AV correlators show clear even-odd oscillations at
both temperatures, whereas these staggered artifacts are less clear in
the S/PS correlators. This has clear implications for the fits: the former
always requires a doubled-parity fit of the form given in \eqn{2masfit},
whereas for the latter a single parity form may suffice.

In \tbn{massPS_b5.415} we show that this is indeed correct.  In the
PS channel a single mass fit is acceptable judging by the value of
$\cdof$, and the fitted value does not change significantly when a
doubled-parity fit is performed. In fact, when a doubled-parity fit is
attempted to the data, the mass of the parity partner is ill-determined.
In the V channel, on the other hand, the doubled-parity fit turns out to
be indispensable.  In the AV channel the behaviour is similar to that in
the V.  In the S channel a single mass suffices, although one needs a
$(-1)^z$ factor multiplying the exponential to take care of staggered
oscillations.  Only at the highest temperature were we able to extract
a second mass from the S/PS correlators. The table also shows that the
fitted masses are reasonably stable against changes in the fit range in
both the PS and V channels. We find similar results for all $T$.

Interestingly, although the covariance matrix is nearly singular (the
smallest correlation coefficient being about 0.8), the difference between
the parameters extracted using or neglecting the covariance matrix in
the PS channel is marginal. The major difference seems to be that the
value of $\chi^2$ obtained when covariances are neglected are clearly
too small for the usual statistical interpretation.

\bef[!tbh]
\begin{center}
\includegraphics[scale=0.7]{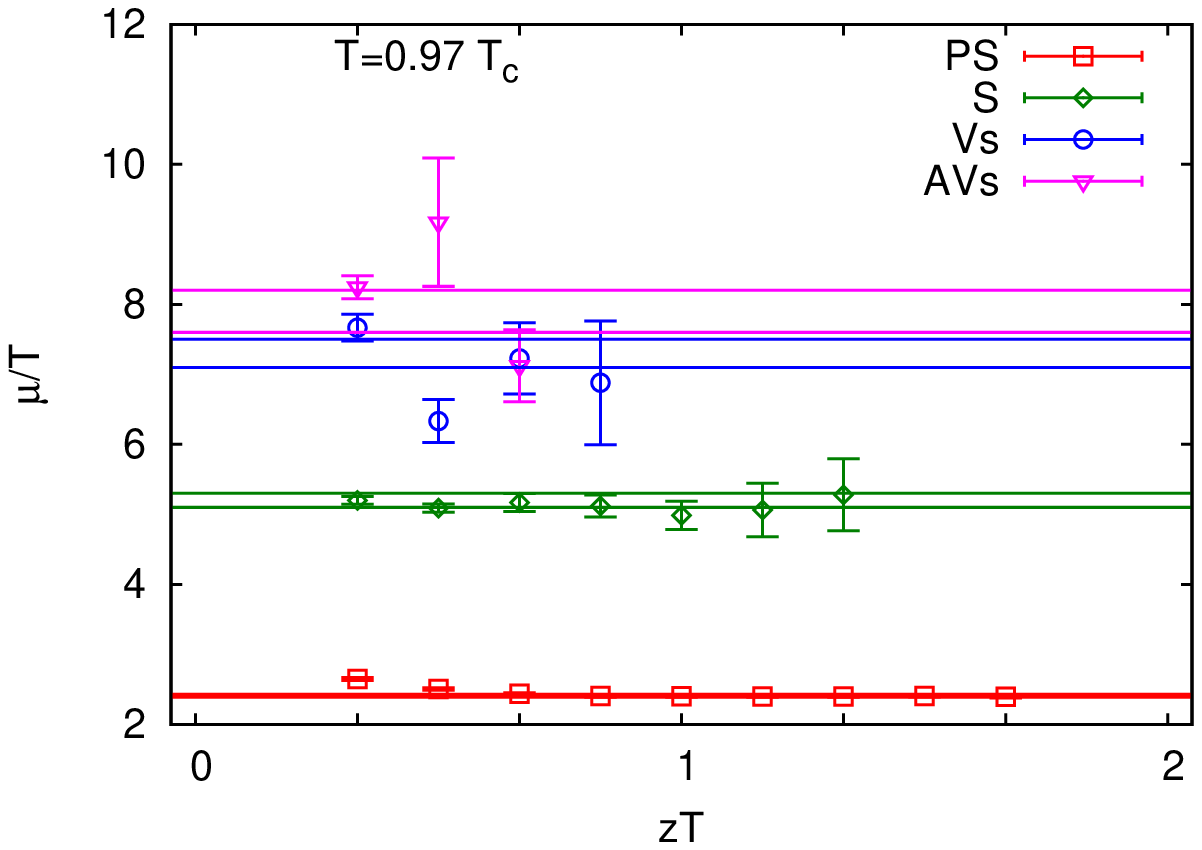}
\includegraphics[scale=0.7]{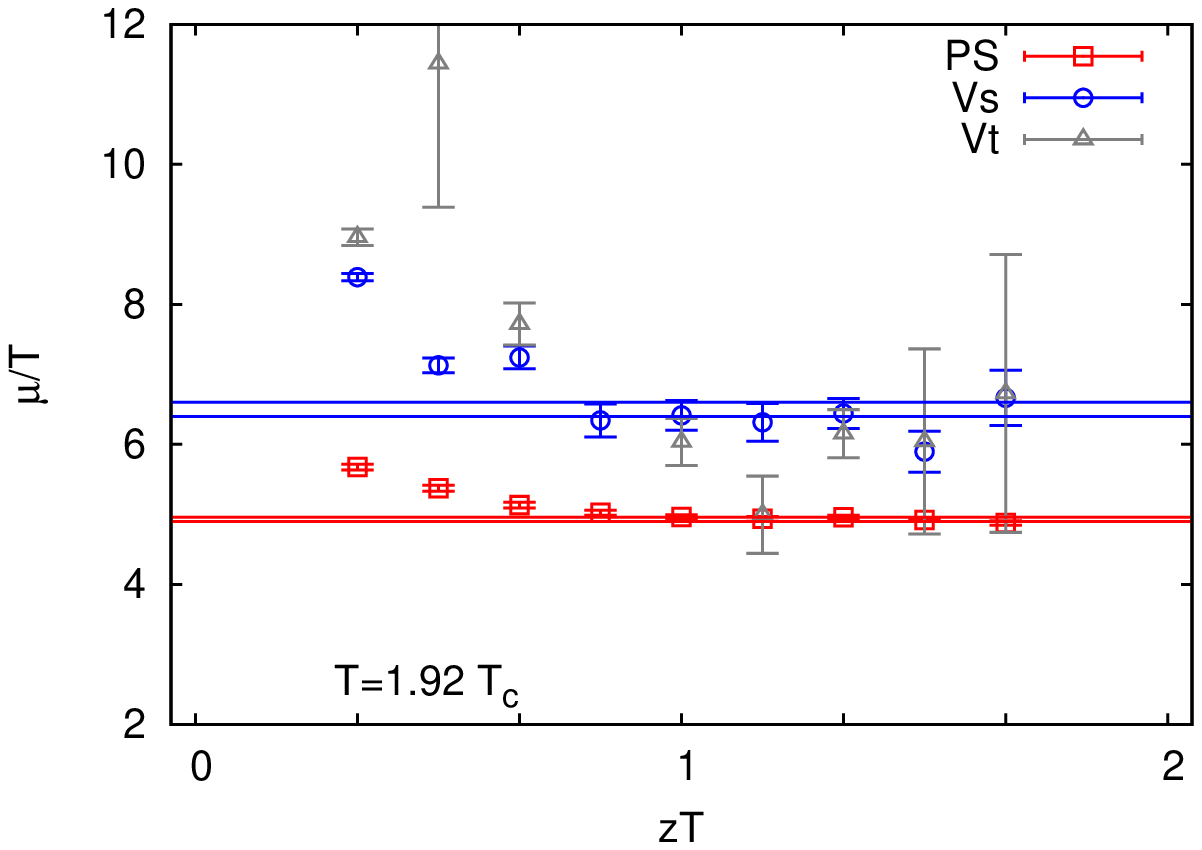}
\end{center}
\caption{The extracted local masses for the mesons at representative
temperatures in the two phases on a lattice with $\zeta=4$. In the high
temperature phase, where chiral symmetry is restored, only the PS, Vs
and Vt masses are shown since the S, AVs and AVt exactly coincide.}
\label{fig:localm}\eef

In \fgn{localm} we demonstrate that the fitted masses agree
with the local mass extracted from eq.\ (\ref{localm}). We draw attention
to the fact that the local masses exhibit a very well developed plateau,
indicating that the correlation functions of fixed parity can be well
described by a single mass. The spatial structure of chiral symmetry
restoration shown in \fgn{chproj3} is visible also in the local masses at
high temperature.

\bet[!tbh]
 \begin{center}
\begin{tabular}{c|c|c|c|c|c|c|c}
\hline
$T/T_c$ &        & PS      & S       & Vs    &   AVs    &   Vt    &  AVt   \\
\hline
1.92 & $\mu_1/T$ & 4.93(3) & 5.1(6)  & 6.5(1)  & 12.6(4)&  9.4(3) & 6.1(2) \\ 
     & $\mu_2/T$ & 5.1(7)  & 4.93(3) & 12.8(4) &  6.5(1)&  6.1(2) & 9.3(4) \\
$\chi^2$     &   & 15.8    & 16      & 9.2     &   9.3  &   7.4   & 7.2    \\  
\hline
1.48 & $\mu_1/T$ & 4.44(4) & 17(1)   & 6.8(1)  & 5.8(4) &  8.9(4) & 6.6(4) \\
     & $\mu_2/T$ & ---     & 4.5(4)  & 5.6(4)  & 6.8(1) &  6.6(4) & 8.7(4) \\
$\chi^2$ &       & 8.1     & 1.8     & 5.2     & 5.3    &  9.6    & 9.2 \\
\hline
1.33 & $\mu_1/T$ & 4.15(2) & 7.0(7)  & 6.7(2)  & 9.6(6) &  8.7(3) & 6.8(5) \\
     & $\mu_2/T$ & ---     & 4.22(4) & 8.7(5)  & 6.8(2) &  6.6(6) & 8.9(4) \\
$\chi^2$ &       & 31.8    & 9.2     & 11.4    & 11.3   &  6.0    & 7.7 \\
\hline
1.21 & $\mu_1/T$ & 3.31(7) & ---     & 6.5(2)  & 9.4(8) &  8.1(4)  & 6.1(8) \\
     & $\mu_2/T$ & ---     & 3.91(5) & 15.0(6) & 6.8(2) &  6.8(8)  & 9.1(4) \\
$\chi^2$ &       & 18.2    & 6.3     & 5.4     &  12.5  &  2.2     & 3.8 \\
\hline
1.012& $\mu_1/T$ & 2.65(4) & 23(3)   & 6.9(3)  & 4.5(9) &  6.4(5)  & 5.7(7) \\
     & $\mu_2/T$ & ---     & 4.3(1)  & 4(2)    & 8.8(4) &  5(1)    & 5.3(6) \\
$\chi^2$ &       & 21.2    & 7.4     & 5.2     & 2.1    &  1.0     & 2.8  \\
\hline
1.00 & $\mu_1/T$ & 2.54(3) & ---     & 5.9(3)  & 5.2(6) & 7.9(4)   & 13(2) \\
     & $\mu_2/T$ & ---     & 4.6(2)  & 9.8(9)  & 8.6(4) & ---      & 11.6(5) \\
$\chi^2$ &       & 11.9    & 7.0     & 6.1     & 4.6    &  3.2     & 7.8  \\
\hline
        &        & PS      & S       & V     &   AV     &         &        \\
\hline
0.99 & $\mu_1/T$ & 2.47(2) & 5(1)    & 5.5(2)  & 6.4(5) &          &       \\
     & $\mu_2/T$ & ---     & 4.6(2)  & 12(1)   & 6.0(4) &          &        \\
$\chi^2$ &       & 8.4     & 2.6     & 6.3     & 4.3    &          &       \\
\hline
0.97 & $\mu_1/T$ & 2.41(2) & 4(1)    & 7.0(2)  & 7.0(3) &          &        \\
     & $\mu_2/T$ & ---     & 5.2(1)  & 11(1)   & 7.7(3) &          &        \\
$\chi^2$ &       & 11.0    & 3.6     & 6.5     & 1.7    &          &      \\
\hline
0.94 & $\mu_1/T$ & 2.35(2) & 4(2)    & 5.9(2)  & 4(1)   &          &        \\
     & $\mu_2/T$ & ---     & 5.2(2)  & ---     & 14.4(3)&          &        \\
$\chi^2$ &       & 15.3    & 3.2     & 2.8     & 2.4    &          &      \\
\hline
0.92 & $\mu_1/T$ & 2.31(2) & 5(1)    & 7.3(3)  & 7.3(7) &          &      \\
     & $\mu_2/T$ & ---     & 5.7(2)  & ---     & 10.0(7)&          &         \\
$\chi^2$ &       & 7.5     & 1.3     & 6.5     & 7.6     &          &       \\
\hline
0.89 & $\mu_1/T$ & 2.27(2) & ---     & 5.5(6)  & 7(1)   &         &      \\
     & $\mu_2/T$ & ---     & 4.9(4)  & 7(2)    & 5(1)   &          &      \\
$\chi^2$ &       & 5.6     & 7.7     & 4.3     & 12.2   &          &      \\ 
\hline
\end{tabular}
\end{center}
\caption{Screening masses at different temperatures. The fit range was
$z/a$=3--11 except for the S/PS at $1.92T_c$, where the fit range was
$z/a$=4--11 (the larger range gave very large $\chi^2$ without appreciably
changing the best fit values). Also note that except at $1.92 T_c$, the PS
fit was done with a single-parity fit form, and therefore has two more degrees
of freedom than the other channels. A dash indicates that some mass could not
be obtained because staggered oscillations were not visible. At temperatures
below $T_c$ the analysis was performed on the V/AV channels.}
\label{table:scrnmas}\eet

\bef[!tbh]
\begin{center}\includegraphics{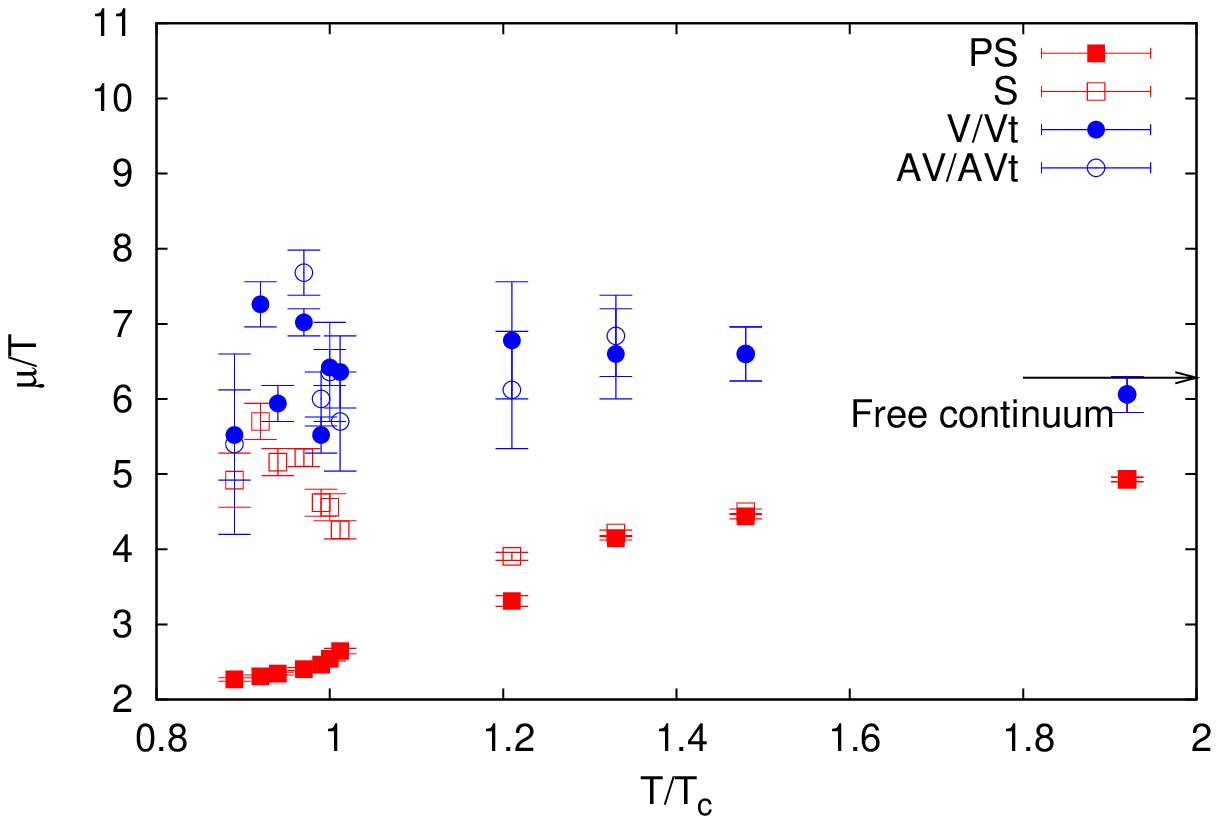}\end{center}
\caption{Screening masses as functions of $T/T_c$. Below $T_c$ the S, PS,
V and AV channels are shown. At $T_c$ and above the S, PS, Vt and AVt are
shown. The free continuum value of 2$\pi$ is indicated with an arrow.}
\label{fig:allmass}\eef

In \tbn{scrnmas} we collect the results for the fitted masses at all
temperatures. We checked in all cases that the local masses for $z>1/T$
were compatible with these fits. For the V/AV correlators we also
checked that if the fits were restricted to $z\le1/T$ the fit results
were generally different. Wherever doubled-parity fits are available,
one can look for chiral symmetry restoration by checking whether or
not the relations of \eqn{csrest} are satisfied. Consistent with the
analysis of \scn{csb}, we find that this happens only for $T\ge1.33T_c$
in the S/PS channels. Surprisingly, the equalities of \eqn{csrest} hold
in the V/AV channels, within statistical errors, from just above $T_c$.

In \fgn{allmass} we plot the lowest screening mass in each channel
as a function of $T/T_c$. Above $T_c$ we could plot
three channels. To avoid clutter we plotted only the S/PS and Vt/AVt
channels. As one sees in \tbn{scrnmas}, the lowest Vs/AVs masses are
slightly larger, but consistent with Vt/AVt at the 2-$\sigma$ level. All
the features discussed are clearly visible here. Also visible is the fact
that $\mu_{PS}/T$ increases monotonically with $T$ whereas $\mu_S/T$
dips near $T_c$. Note also that $\mu_{Vt}/T$ may approach its ideal
gas value from above, becoming consistent with the limit already at
$T\simeq2T_c$. However $\mu_S/T$ remains about 20\% below this limit
even at the highest temperature we explored. We shall return to this
point later when we discuss the continuum limit.

\subsection{The role of explicit chiral symmetry breaking}\label{sec:xcsb}

\bef[tbp]
\begin{center}
\includegraphics[scale=0.7]{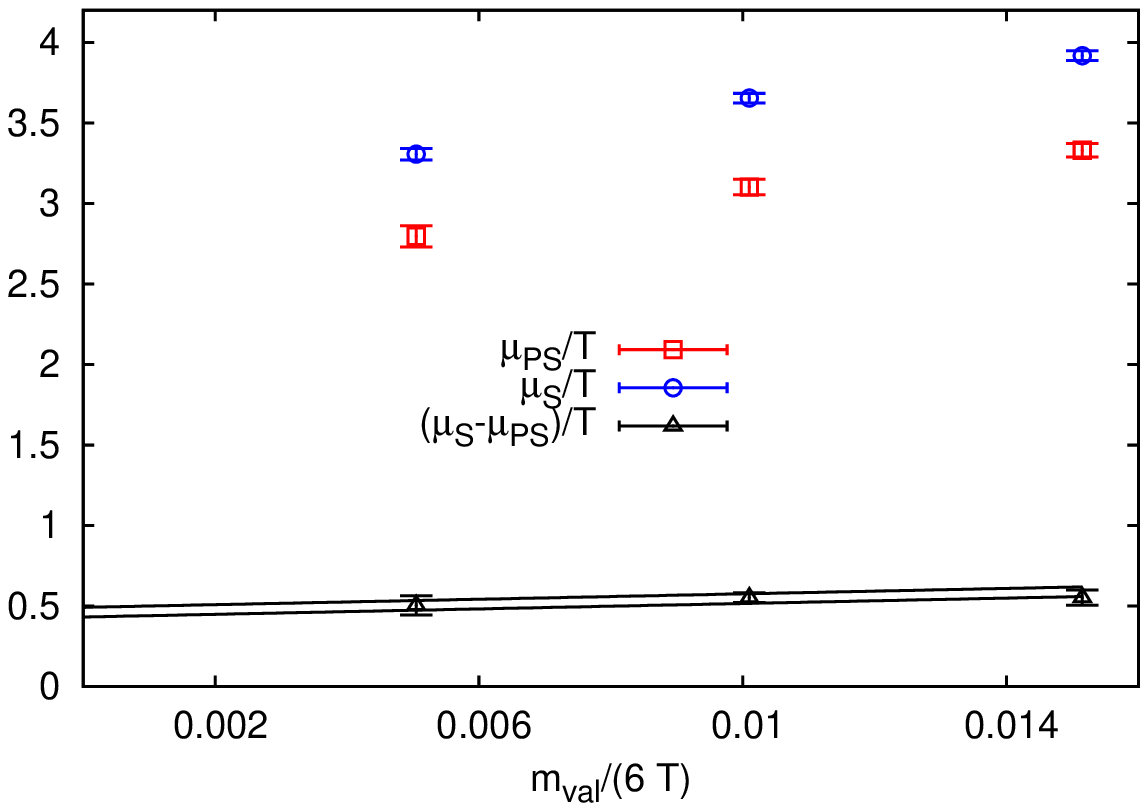}
\includegraphics[scale=0.7]{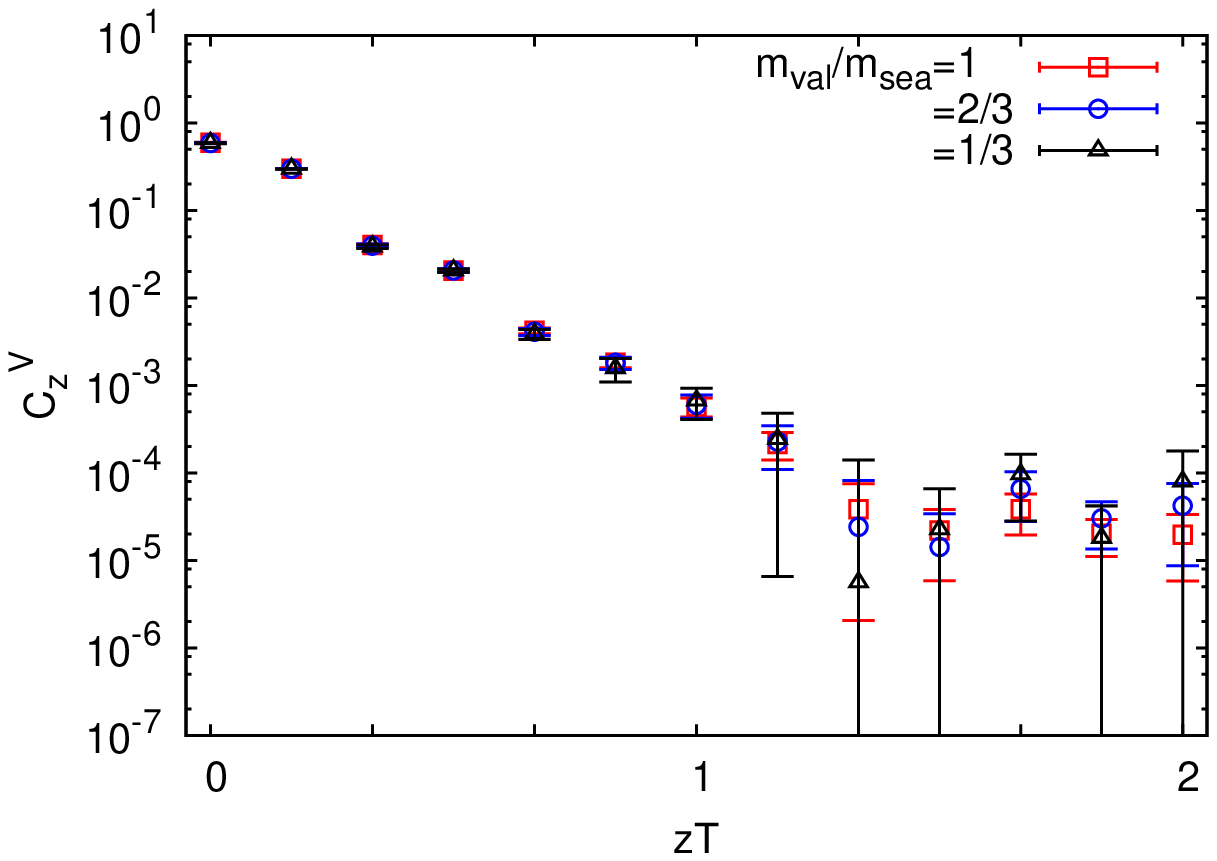}
\end{center}
\caption{The valence quark mass dependence of various quantities at
$T=1.21T_c$ when the bare sea quark mass is held fixed at the value
$m/T_c=0.1$ (corresponding to $m_\pi\simeq230$ MeV). The screening masses 
in the S/PS sector depend significantly on the valence quark mass,
whereas the V sector screening correlators are insensitive. }
\label{fig:mqmas}\eef

In free field theory one has no pion and the explicit chiral symmetry breaking
scale is the quark mass. In this case one has
\beq
   \frac\mu T = 2\sqrt{\pi^2 + \left(\frac mT\right)^2}
     \simeq 2\pi\left[1+\frac12\left(\frac m{\pi T}\right)^2\right].
\label{xcsb}\eeq
Substituting the bare quark mass into this expression, it can be seen
that the effect is of the order of a few parts in $10^5$, and hence
negligible. However, it turns out that in weak-coupling theory one has to
insert in the above equation the thermal mass of the quark \cite{htl}.
Since this is $gT/\sqrt3$, and $g$ is large near $T_c$ \cite{precise},
the effect can be significant. Of course, when $g$ is large, the weak-coupling
theory is unlikely to be quantitatively useful, and should be taken only
as an indication that one must explore the quark mass dependence of the
chiral symmetry breaking seen in the screening masses.

\bef[tbp]
\begin{center}\includegraphics{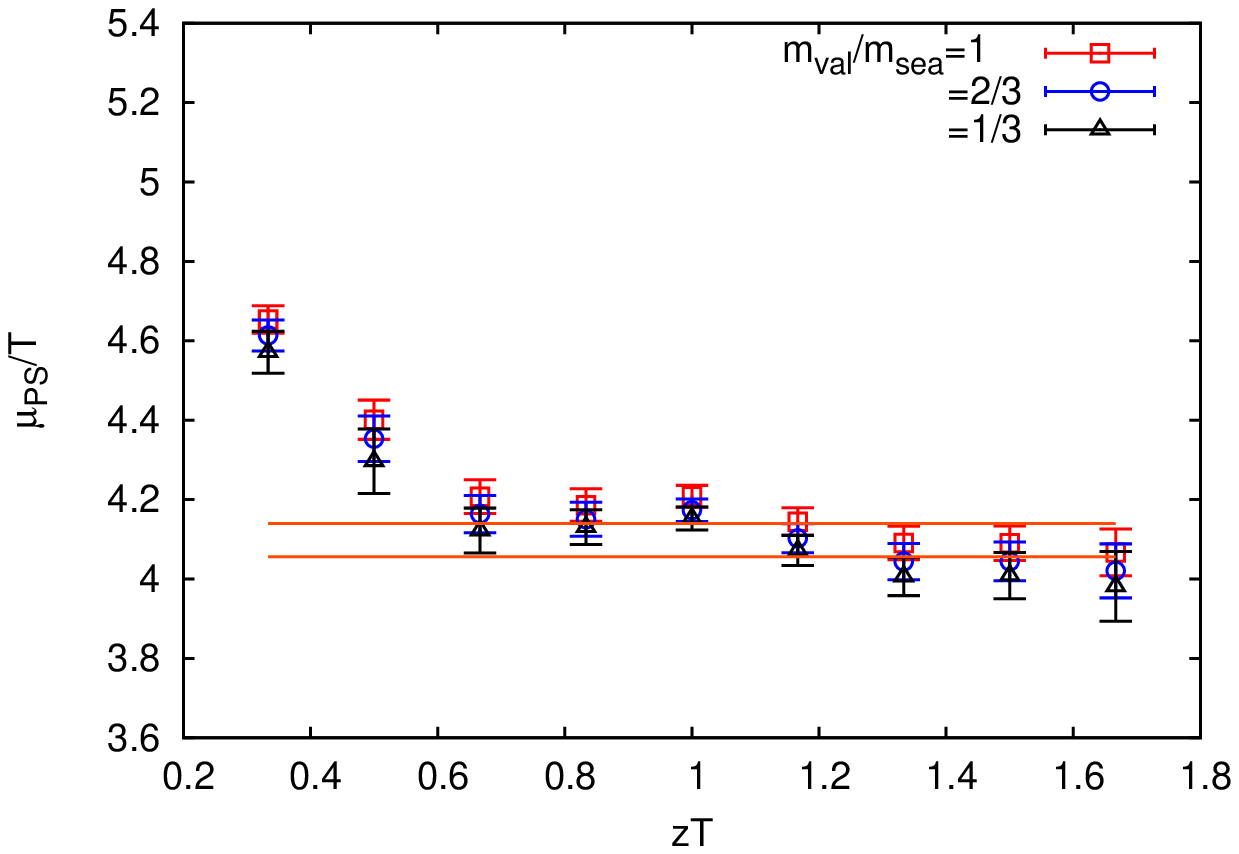}\end{center}
\caption{Local masses in the PS channel for different valence quark masses
at $T=1.33T_c$ with $\zeta=4$. The horizontal lines are the 1-sigma bounds
from the fit to the correlator for $m_{val}=m_{sea}$.}
\label{fig:mqmps}\eef

Changing the quark mass involves in principle a completely new set of
computations, and lies beyond the scope of this paper. However, it is
possible to change the valence quark mass ($m_{val}$) without changing
the sea quark mass ($m_{sea}$). As a result, such a restricted study can
be done without a large cost in CPU time. In view of this, we repeated
our analysis above $T_c$ with $m_{val}$ chosen to be 2/3 and
1/3 of $m_{sea}$. Sample results are shown in \fgn{mqmas}.

We find that a change in the valence quark mass has insignificant effect
in the V/AV channels (see, for example, the second panel of
\fgn{mqmas}). However, there are statistically significant changes in
the S/PS sector. Both the S and PS screening mass increase with $m_{val}$.
The difference also increases, although the limiting value for $m_{val}=0$
is finite. We find that
\beq
   \left.\frac{\mu_S-\mu_{PS}}T\right|_{T/T_c=1.21} = 0.46\pm0.03 \qquad
    (\mathrm{for}\; m_{sea}/T_c=0.1,\; m_{val}=0).
\label{extrapolation}\eeq
It is not inconceivable that decreasing $m_{sea}$ in a future computation
can lower this mass difference further. Note, however, that $\mu_{PS}/T$
and $\mu_S/T$ both drop as the quark mass changes, and move further away
from the weak-coupling expectation. 

In fact, the dependence of correlation functions on the bare quark mass can
be deeper than what was discussed above. We illustrate this in
\fgn{mqmps}, where local masses in the PS channel are shown for
several values of $m_{val}$. It seems that with decreasing $m_{val}$ the
plateau in the local masses becomes less well developed; for the smallest
$m_{val}$, in fact, a distinct slope is visible.

\subsection{The effect of finite volume}\label{sec:fss}

\bet[!tbh]
\begin{center}
\begin{tabular}{|c|c|c|c|c|c|}
\hline
$\zeta$& & PS & S & V & AV \\
\hline
4/3&  $\mu_1$ & 2.45(4) & ---    & 8.0(2) & 4.3(6) \\
   &  $\mu_2$ &  ---    & 5.0(3) &  ---   & 7.7(6) \\
 2 &  $\mu_1$ & 2.41(2) & ---    & 6.8(3) & 13.1(7)\\
   &  $\mu_2$ & ---     & 5.5(2) & 14.7(5)& 9.2(5) \\
 3 &  $\mu_1$ & 2.44(3) & 6(1)   & 6.5(2) & 8.0(3) \\
   &  $\mu_2$ & ---     & 5.3(2) & 6.4(8) & 9.1(3) \\
 4 &  $\mu_1$ & 2.35(2) & 4(2)   & 6.4(3) & 7.6(3) \\
   &  $\mu_2$ &  ---    & 5.2(2) & 29(4)  & 9.4(2) \\
 5 &  $\mu_1$ & 2.39(2) & 4.2(-) & 6.6(4) & 8.1(5) \\
   &  $\mu_2$ & ---     & 5.5(4) & 9(1)   & 8.6(4) \\
\hline
\end{tabular}
\end{center}
\caption{Screening mass estimates at $T=0.94T_c$ at fixed $a=1/(6T)$ with
changing spatial volume $(\zeta/T)^3$.}
\label{table:massV}\eet

\bef[!tbh]
\begin{center}\includegraphics{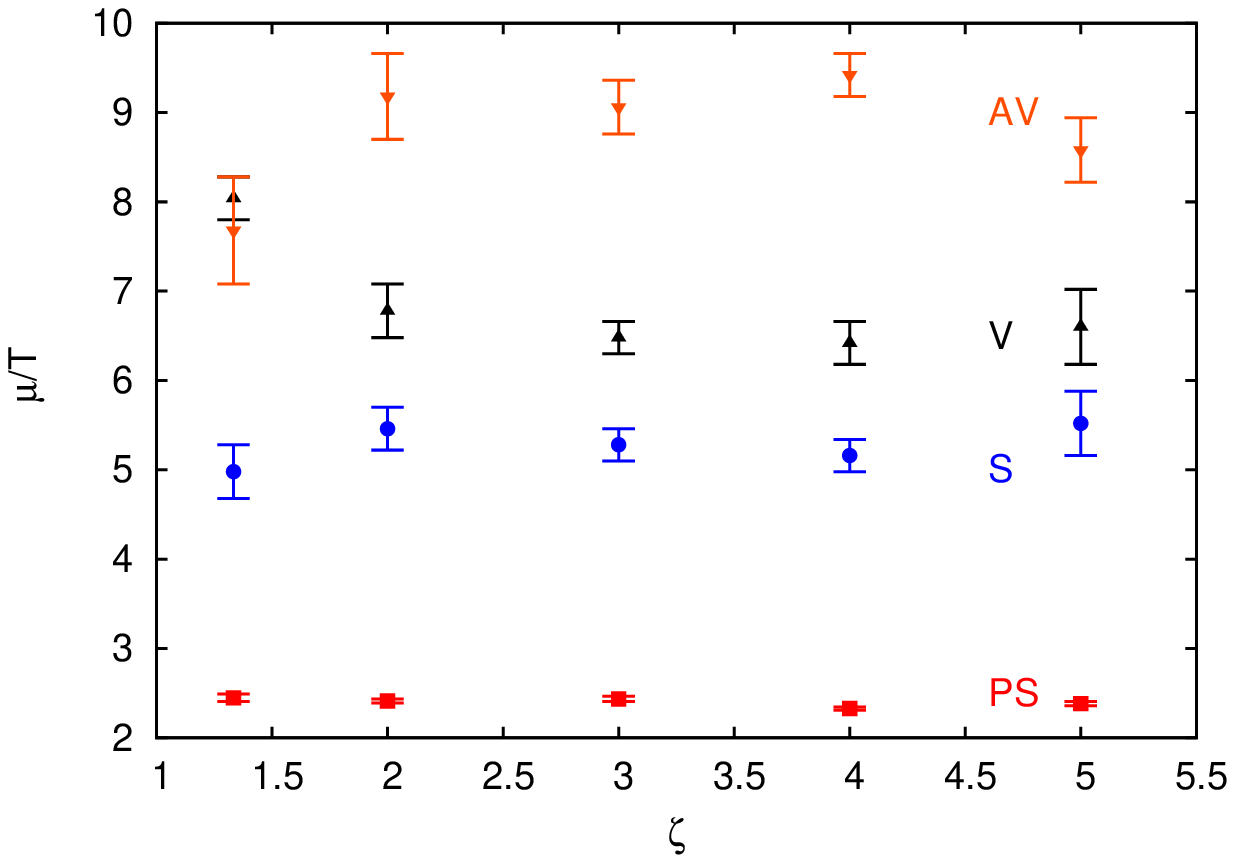}\end{center}
\caption{Screening masses in various channels as a function of $\zeta$ at
$T=0.94 T_c$. Note that there is nothing special about the S channel
masses.}
\label{fig:massV}\eef

Finite volume effects in the high temperature phase of QCD have been
explored earlier \cite{rvgsgpm} and are by now well understood. These
effects are under control as long as $\mu_{PS}$ is large compared to the
inverse size of the smallest lattice dimension, \ie, $T$. Since we find
$\mu_{PS}/T>2.5$ in the high temperature phase, we expect that finite
volume effects are under control for $T>T_c$.

In this paper we report on the magnitude of finite volume effects,
below $T_c$. We studied the screening correlators on lattices with
$4/3\le\zeta\le5$ at $T=0.94T_c$. As shown in \tbn{massV} and \fgn{massV},
finite volume effects are invisible within statistical errors. Again,
since $\mu_{PS}/T\simeq2.3$, this may not seem unexpected.

It is worth the remark that while such studies can have some bearing
on decay widths at finite temperature, much larger lattices and smaller
quark masses and lattice spacings may be required for those. For example,
the scalar under study cannot decay into two pions, and must have at
least three pions in the final state. This is ruled out kinematically
at present.

On the other hand, since $\mu_S/\mu_{PS}\simeq2.2$, one must ask
whether the long-distance behaviour of the S correlator is determined
by a single scalar exchange or multi-particle exchanges (this is the
finite temperature analogue of particle decays, and we shall save space
by using the word decay). In the continuum theory, this non-isosinglet
scalar cannot decay into two pions.  From this point of view, since
$\mu_S/\mu_{PS}\simeq2.2<3$, one could expect that the scalar does not
decay. However, with staggered quarks on a lattice, there are spurious
two pion states (taste multiplets) through which the scalar current
could be correlated \cite{scalar}. The featureless behaviour of the
volume dependence of the screening masses implies that no decays occur.
However, this is a weak statement, because a 7\% taste symmetry breaking
is sufficient to forbid this decay on the largest lattice which we used.

\subsection{The continuum limit}\label{sec:conti}

\bet[!tbh]
\begin{center}
\begin{tabular}{c|c|c|c|c}
\hline
$T/T_c$ & \multicolumn{2}{c}{$a=1/(4T)$} & \multicolumn{2}{|c}{$a=1/(6T)$} \\
\hline
 & S/PS & Vt/AVt & S/PS & Vt/AVt \\
\hline
1.5 & $3.67\pm0.02$ & $5.44\pm0.08$ & $4.44\pm0.04$ & $6.6\pm0.4$ \\
2.0 & $4.08\pm0.01$ & $5.72\pm0.04$ & $4.93\pm0.03$ & $6.1\pm0.2$ \\
\hline
\end{tabular}
\end{center}
\caption{The lattice spacing dependence of $\mu/T$ in various quantum
number channels. The temperature scale is rounded off. Data for $a=1/(4T)$
is taken from \cite{rvgsgpm}.}
\label{table:conti}\eet

In \tbn{conti} we extract the values of the screening masses in
units of temperature, $\mu/T$ for two different temperatures in the chirally
symmetric phase, at two different lattice spacings. From two pieces of data
in each case it is hard to construct a continuum extrapolation. However, we
can test whether the extrapolation is consistent with the expectation $\mu\simeq2\pi T$ by attempting a fit to the form
\beq
   \left.\frac\mu T\right|_{N_t} = 2\pi + \frac{s}{N_t^2}.
\label{conti}\eeq
It is possible to get reasonable fits in the V/AV channel, yielding
\beq
   s = \begin{cases}-13\pm2 & (T=1.5T_c)\\ -9.0\pm0.2 & (T=2T_c)\end{cases}
\label{results}\eeq
This is consistent with previous results. However, in the S/PS case this
procedure fails to yield the expected result; the extrapolated screening length
remains below the ideal gas value. One cannot rule out the possibility that
the weak-coupling result emerges at even smaller lattice spacings.

We note however, that the screening of meson-like correlations in the weak
coupling theory occurs, not through the exchange of a single particle,
but through multiparticle exchange. As a result, the zero-momentum
correlator is not expected to be strictly exponential, but to have some
curvature. Such a curvature was actually seen in computations in the
quenched theory using staggered \cite{rvgsgq} and Wilson \cite{wissel}
quarks using much smaller lattice spacings. An easily observed feature
of such a curvature is that local masses do not show a plateau, but
change continuously with $z$. Such behaviour was neither seen here 
(see \fgn{localm}), nor with p4 improved quarks at the same
renormalized quark mass \cite{hotqcd}. However, as pointed out in
\scn{xcsb}, when the quark mass is lowered such a feature
could emerge.

\section{Summary}\label{sec:summ}

We studied screening correlation functions and screening masses
of meson-like probes of strongly interacting matter, both in the
low temperature hadron and high temperature plasma phases. We used
configurations described in \cite{rvgsgNt6}; these were generated using
two flavours of dynamical staggered quarks with masses tuned to give
$m_\pi\simeq230$ MeV. Most of the results come from lattice spacing
$a=1/(6T)$, although we have attempted to check assumptions about the
continuum limit using earlier measurements with the same renormalized
quark mass and lattice spacing $a=1/(4T)$. We have explicitly checked
for finite volume effects, and found that these are negligible when 
we take the aspect ratio $\zeta=4$.

We checked that the correlators at $T<T_c$ effectively have the
symmetries of the $T=0$ transfer matrix, and that there is a fairly abrupt
transition at $T_c$ to the symmetries of the screening transfer matrix
(see \fgn{vector}).  Although the QCD cross over occurs at $T_c$, we found
a lack of parity doubling in the spectrum of screening masses up to a
temperature of $1.33T_c$ (see, for example, \fgn{sus1} and \tbn{zero}).
We argued that explicit chiral symmetry breaking due to the quark mass
is visible up to a temperature of $1.35T_c$, as seen here. This argument
leads us to believe that in the limit of realistic quark masses, such a
breaking would not be visible above $T_c$.  Interestingly, there turns
out to be a lot of structure in this apparent breaking of chiral symmetry
above $T_c$. The V/AV correlators are equal to each other at distances
$z>1/T$, and the chiral symmetry breaking in this sector is entirely
a short distance effect (see \fgn{chproj3}). In the S/PS sector the
non-degeneracy of the correlators persists into the long distance regime.

Non-degeneracy of the S/PS correlators could also be due to $U_A(1)$
symmetry breaking. This is suspected to persist well into the high
temperature plasma phase \cite{rvgsgrl0}. We tested what happens to
the S/PS difference as the valence quark mass is changed (see
\fgn{mqmas}).  Our results imply that other physics effects
can be disentangled from the explicit symmetry breaking effect due
to finite quark mass only through computations with smaller quark
masses. Alternatively, one could examine the short distance part of the
V/AV correlators for signals of such microscopic physics.

We made the first study of hadron decays at finite temperature (below
$T_c$) through a systematic exploration of the volume dependence of
screening masses. We found no significant volume dependence (see, for
example, \tbn{massV} and \fgn{massV}), indicating the stability of the
scalar. As we discussed already, this study needs to be carried out with
smaller quark masses so that $\mu_S/\mu_{PS}>3$, or at smaller lattice
spacings, so that taste violations are reduced.

We combined the analysis of this paper with data from an earlier source
\cite{rvgsgpm} in which the same renormalized quark masses were used
to study screening at a coarser lattice spacing, $a=1/(4T)$ to explore
the continuum limit. Being restricted to only two values of the lattice
spacing at each $T$, we ask whether the continuum limit of screening
masses is compatible with the ideal gas expectation, $2\pi T$, in the
high-temperature phase. We find that it is, in the V/AV channels, but
not in the S/PS channels.

If the high temperature phase is deconfined, then correlations of static
currents with meson quantum numbers must be mediated by the exchange of
a quark anti-quark pair. The most straightforward signal of this is that
the local masses do not show a well-developed plateau. In most of our
studies we did not see this. Only in a study with rather small valence
quark masses did we see a signal of such behaviour (see \fgn{mqmps}).
Studies with lower sea quark masses in the future will be needed to
resolve the question of deconfinement above $T_c$ in QCD with physical
quark masses.

\section*{Acknowledgments}

The computations were performed on the CRAY X1 of the Indian Lattice
Gauge Theory Initiative (ILGTI) in TIFR, Mumbai. We thank Ajay Salve
and Kapil Ghadiali for technical support. DB wishes to acknowledge
useful discussions with Saumen Datta, Nilmani Mathur and Jyotirmoy Maiti.
\appendix

\section{Covariance matrices and statistical data compression}\label{sec:fitstat}

\bef[!tbh]
\begin{center}
\includegraphics[scale=0.7]{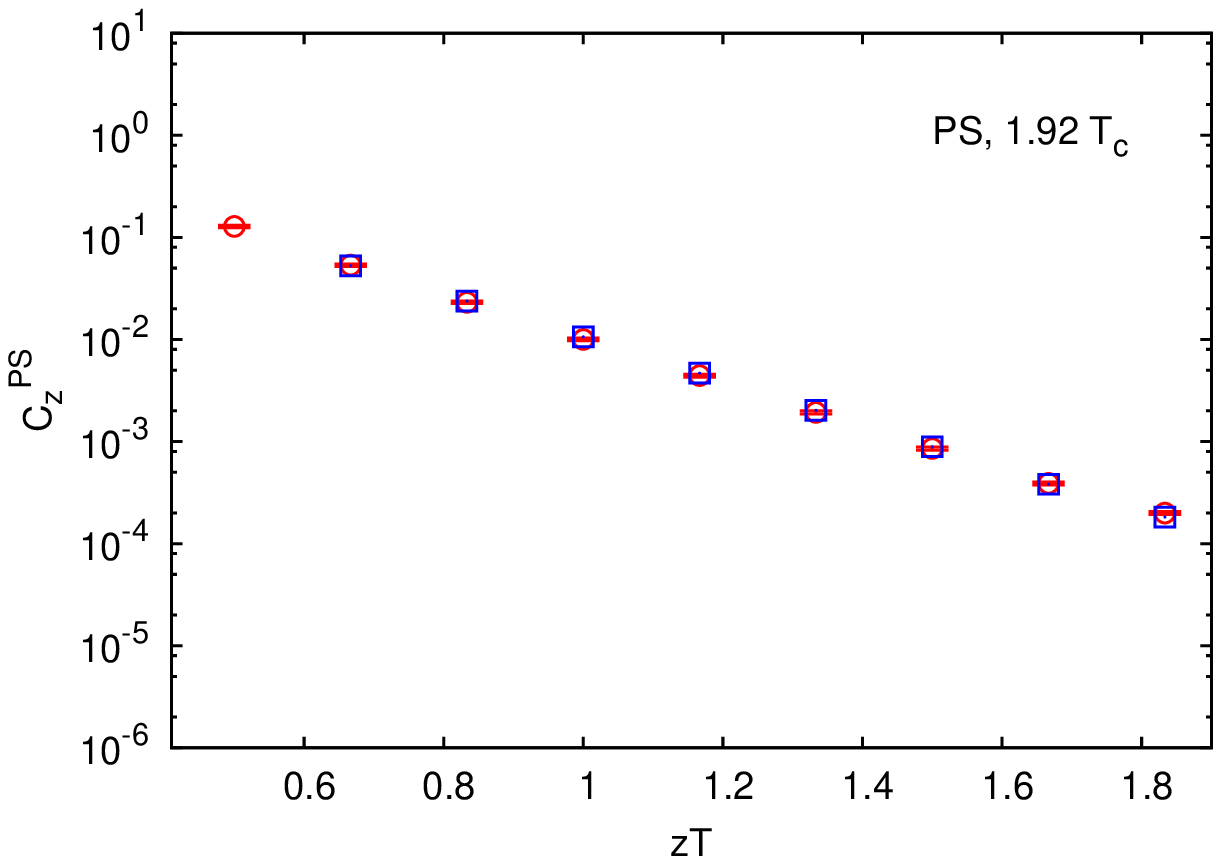}
\includegraphics[scale=0.7]{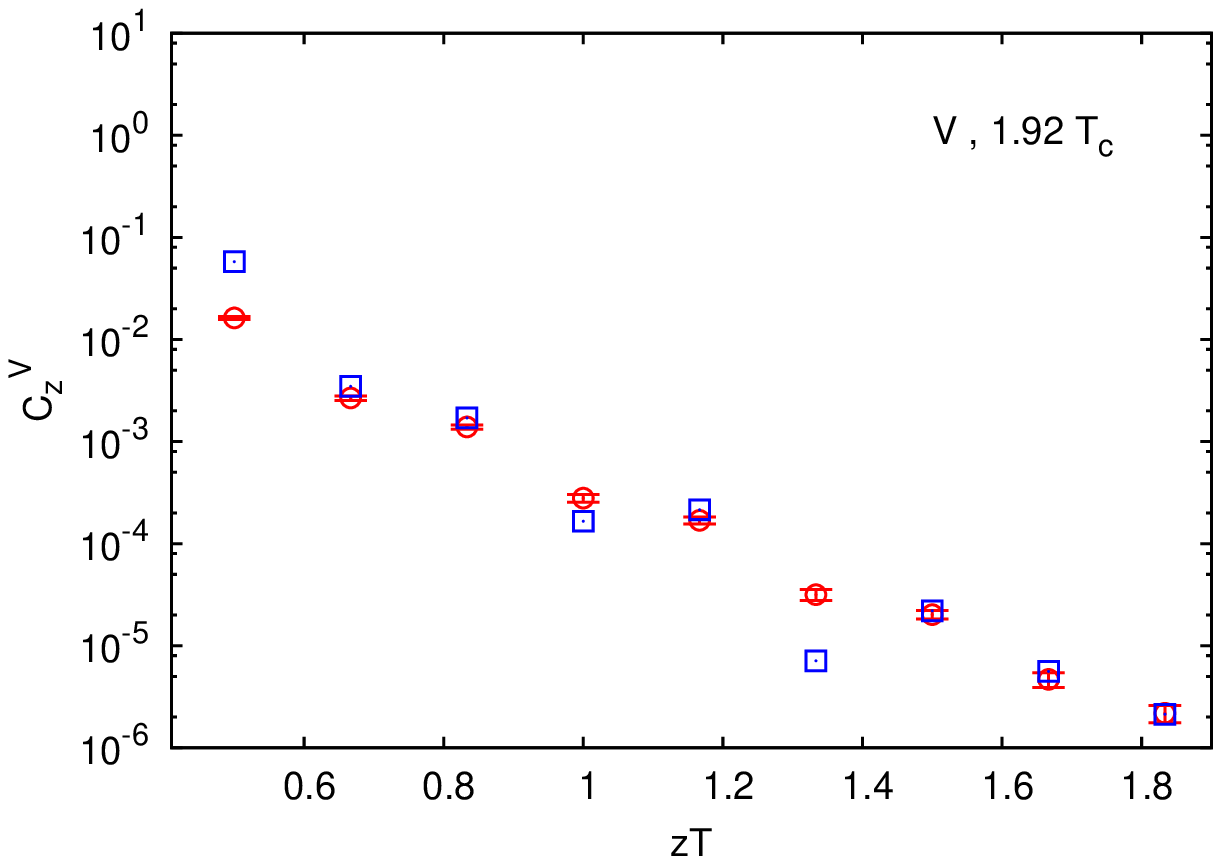}
\includegraphics[scale=0.7]{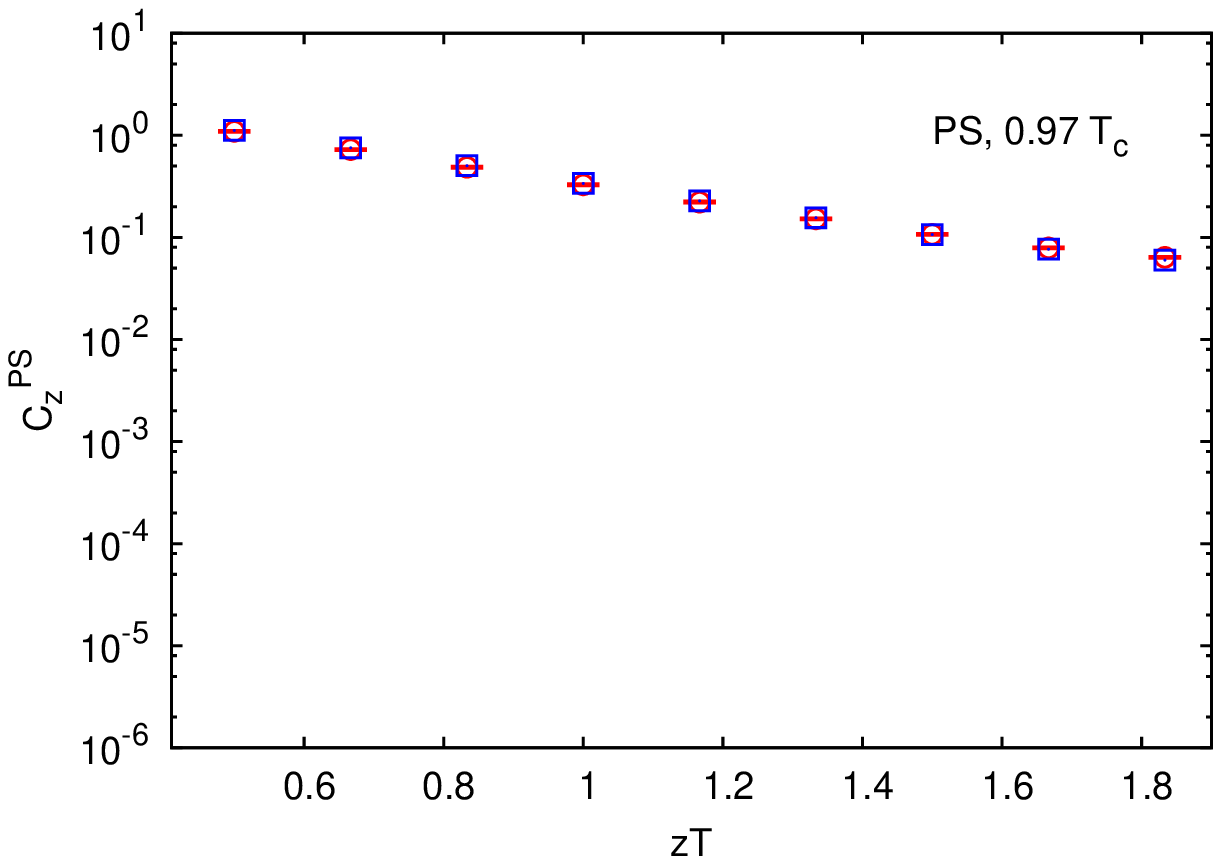}
\includegraphics[scale=0.7]{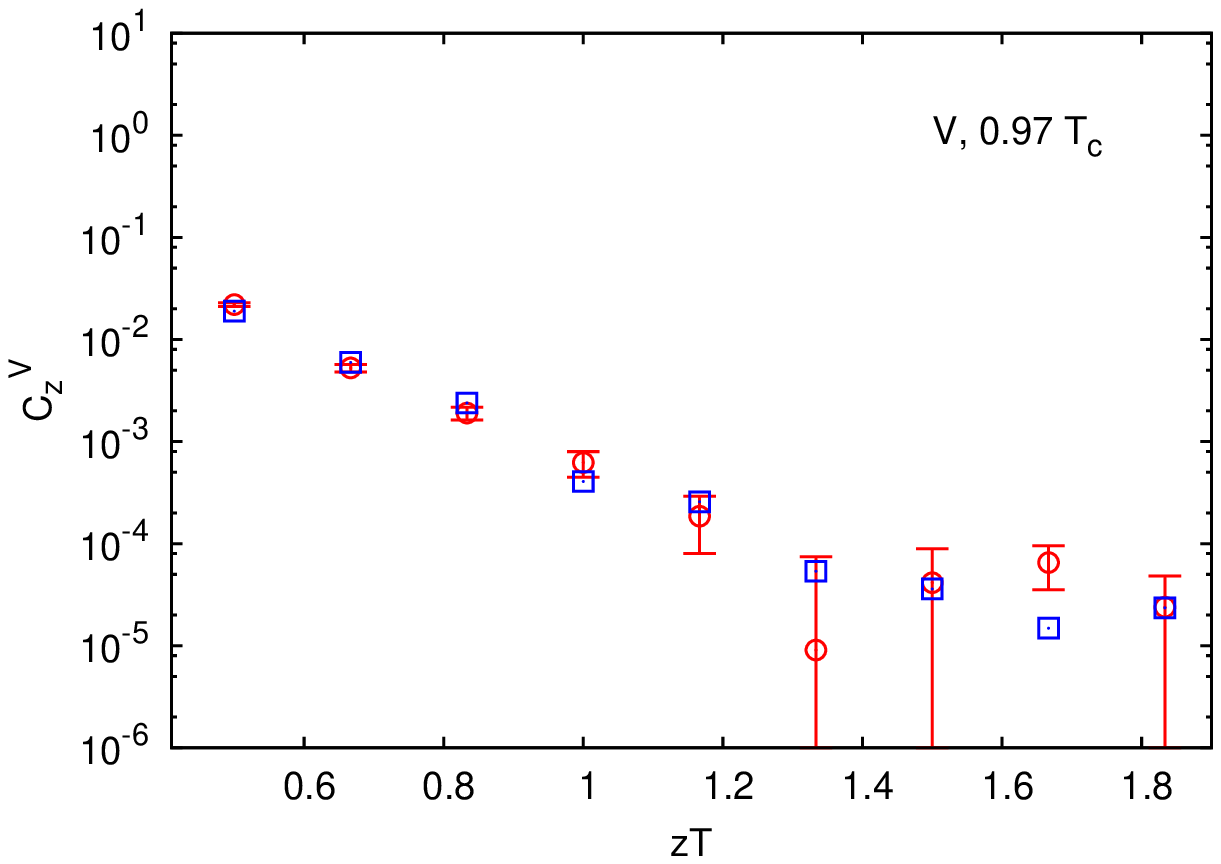}
\end{center}
\caption{The measured correlator (red circles) and the compressed version
obtained by projection on to a single eigenvector of $\Sigma$ (blue squares)
for the PS and V at $T=1.92T_c$ and $0.97T_c$.}
\label{fig:eigen}\eef

The correlation functions are evaluated on a fixed set of gauge
configurations. In general, the estimates of the correlator at different
separations are not statistically independent. The covariance matrix of
the correlators measures the degree of independence, and is defined as
\beq
  \Sigma_{zz'} = \left\langle\left(C^\gamma_z-\langle C^\gamma_z\rangle\right)
     \left(C^\gamma_{z'}-\langle C^\gamma_{z'}\rangle\right)\right\rangle,
\label{cov}\eeq
where the averages are over the ensemble of gauge configurations. The
diagonal elements are the variances. The matrix is positive and symmetric
by construction, so that it has well defined (positive) eigenvalues and
orthogonal eigenvectors. It is found that it often has large condition
number, \ie, the ratio of the largest and smallest eigenvalues is
large. Any fits to the correlation function are made by minimizing the
$\chi^2$-function
\beq
   \chi^2 = D^t \Sigma^{-1} D,\qquad{\rm where}\qquad
       D=C^\gamma - f,
\label{chisq}\eeq
$C^\gamma$ is the vector of measured correlation functions, with
components $C^\gamma_z$, and $f$ is the corresponding
vector made from the function to be fitted.

The covariance matrix is often inverted using a singular value
decomposition, $\Sigma=UDV^T$ where $U$ and $V$ are orthogonal matrices
(in this case $U=V$) and $D$ is diagonal. The inverse matrix is usually
obtained as $\Sigma^{-1}=VD^{-1}U^T$, with the largest components of
$D^{-1}$ set to zero \cite{press}. This procedure is used to prevent
errors arising from finite precision rounding.  For this part of
the analysis we use Mathematica, which allows arbitrary precision
computations. By tuning the precision of the computation from 10 digits
to 40 digits, we checked for control of rounding errors. Our conclusion
is that the rounding errors can be kept under control without setting
large numbers to zero. As a result, in our ensuing investigations we
are assured that the eigenvalues of $\Sigma$ are the inverses of those
of $\Sigma^{-1}$.

The covariance matrix and its eigenvectors are determined essentially
by the statistical properties of the measurement and not directly by the
correlation function. One extreme can be illustrated by the imagined case
where we generate a very large set of decorrelated gauge configurations,
and use a disjoint subset for evaluation of the correlator at each
$z$. Since the measurements at each of the $z$ are then statistically
independent of each other, $\Sigma$ must be diagonal. As a result,
the eigenvectors have support only on a single separation $z$.


Take the
eigenvalues of $\Sigma^{-1}$ to be $\lambda_\alpha$ with corresponding
eigenvectors $v_\alpha$. Each correlation function can be written as a
linear combination of the $v_\alpha$, \ie,
\beq
   C^\gamma = \sum_\alpha c^\gamma_\alpha v_\alpha.
\label{decom}\eeq
Our purpose in doing this is that if some of the $c^\gamma_\alpha$
turn out to be small then one can perform a noise reduction by dropping
the small terms. Since $c^\gamma_\alpha=C^\gamma\cdot v_\alpha$, the
errors on the dot product are clearly given by
\beq
   \sigma^2(c^\gamma_\alpha) = v_{\alpha z} v_{\alpha z'} \Sigma_{zz'}
      = \frac1{\lambda_\alpha}.
\label{err}\eeq
In writing this formula we have taken $v_{\alpha z}$ to have no
errors. This approximation can be removed, if necessary, by a bootstrap
analysis of the errors in $c^\gamma_\alpha$.  If many of the components,
$c^\gamma_\alpha$, are zero within errors, then we can drop them
from the analysis. In this case we say that the correlation function
is compressible. Such a notion corresponds to that of `lossy data
compression' in a variety of contexts, including jpeg image compression
and mp3 audio compression.

In the imagined case which we discussed before, generically all the
$c^\gamma_\alpha$ will be non-zero and the correlator will not be
compressible.  Only if some of the measurements are so noisy that they
are compatible with zero can we drop them from the fits. This example
illustrates the connection of compressibility with that of statistical
independence.

The actual situation we investigated was the usual one where
the correlation function at all separations were measured on all
configurations and the lowest mass was obtained by a fit to the
long-distance correlation function using the cuts mentioned in \scn{scr},
\ie, $3\le z/a\le13$.  Very surprisingly, we found that the PS correlator
can be compressed to exactly one component, corresponding to the
smallest $\lambda_\alpha$ both above and below $T_c$. This means that
the corresponding eigenvector $v_\alpha$ contains the full information
on the correlator. The V correlator can also be compressed down to a
very small number of components, usually one or two.  These surprising
results are shown in \fgn{eigen}. When we examine the correlation
functions at all distances, then again we find a high, but lesser,
degree of compressibility with two or three eigenvectors being needed
for the description of the data.

What is the origin of these extremely strong covariances in the data?
The gauge configurations we used were chosen so that
thermodynamic operators such as the action, Polyakov loops, chiral
condensate, various quark number susceptibilities, are decorrelated
between one configuration and another. Several of these quantities, for
example, those which involve fermions, are highly non-local. In view of
this, autocorrelations between configurations are not the cause of the
high compressibility of the data.

So it must be the correlation function itself which generates
these covariances, with the result that one of the eigenvectors,
$v_\alpha$ contains all the information present in the measured
correlator. Could this be due to the fact that the Dirac propagators
are highly non-local? If so, the baryon correlators must also be highly
compressible, whereas glueball correlators need not be. Or could it be
that the relatively low masses are at the root of the covariance? If so,
neither the baryon nor glueball correlators need be compressible. Since
correlated fits are standard technology in the fitting of masses,
measuring statistical compressibility of correlators is a trivial
extension of standard lattice measurements. It would be interesting to
have more data in future on the compression of correlation functions.

\end{document}